%% file: cks-gaia.tex
\documentclass[preprint2]{aastex6}

\usepackage{epsfig,natbib}
\usepackage{graphicx}
\usepackage{amsmath}
\usepackage{longtable}
\usepackage{xspace}
\usepackage{hyperref}
\usepackage{lipsum}
\input{newcommand_gen.tex} 
\input{values.tex} 
\input{val_stat.tex} 

\shortauthors{Fulton \& Petigura}
\shorttitle{CKS-Gaia}
\begin{document}

\title{The California-Kepler Survey VII. Precise Planet Radii Leveraging Gaia DR2 Reveal the Stellar Mass Dependence of the Planet Radius Gap}
\author{
Benjamin J.\ Fulton\altaffilmark{1,2}
and
Erik A.\ Petigura\altaffilmark{1,3}
}

\altaffiltext{1}{California Institute of Technology, Pasadena, CA 91125, USA}
\altaffiltext{2}{IPAC-NASA Exoplanet Science Institute Pasadena, CA 91125, USA}
\altaffiltext{3}{Hubble Fellow}

\begin{abstract}
The distribution of planet sizes encodes details of planet formation and evolution.  We present the most precise planet size distribution to date based on \Gaia parallaxes, \Kepler photometry, and spectroscopic temperatures from the California-\Kepler Survey.  Previously, we measured stellar radii to 11\% precision using high-resolution spectroscopy; by adding \Gaia astrometry, the errors are now 3\%.  Planet radius measurements are, in turn, improved to 5\% precision.  With a catalog of $\sim$1000 planets with precise properties, we probed in fine detail the gap in the planet size distribution that separates two classes of small planets, rocky super-Earths and gas-dominated sub-Neptunes.  Our previous study and others suggested that the gap may be observationally under-resolved and inherently flat-bottomed, with a band of forbidden planet sizes.  Analysis based on our new catalog refutes this; the gap is partially filled in.  Two other important factors that sculpt the distribution are a planet's orbital distance and its host star mass, both of which are related to a planet's X-ray/UV irradiation history.  For lower mass stars, the bimodal planet distribution shifts to smaller sizes, consistent with smaller stars producing smaller planet cores.  Details of the size distribution including the extent of the `sub-Neptune desert' and the width and slope of the gap support the view that photoevaporation of low-density atmospheres is the dominant evolutionary determinant of the planet size distribution.
\end{abstract}

%

\section{Introduction}
\label{sec:intro}

NASA's prime \Kepler mission (2009--2013; \citealt{Borucki10a, Borucki16}) is continuing to revolutionize our understanding of planetary astrophysics. {\em Kepler's} success flows from its near continuous high precision photometric monitoring of $\sim$150,000 stars over a four year mission. Among many discoveries, the large and homogeneous \Kepler dataset enabled  demographic studies of large numbers of exoplanets as small as Earth (see, e.g., \citealt{Howard12,Fressin13,Petigura13b}). One startling result from these studies is that nearly every Sun-like star has a planet larger than Earth but smaller than Neptune. Given the lack of such planets orbiting the Sun, \Kepler demonstrated that the Solar System is not a typical outcome of planet formation, in at least that one key respect.

Initially, the basic structure of these ubiquitous 1 to 4~\Re planets was unknown.  It was unclear whether these planets were predominately rocky or had substantial gaseous envelopes. Early clues came  from mass measurements of a few tens of \Kepler planets based on the radial velocity (RV) and transit-timing variation (TTV) techniques \citep{Marcy14,Holman10}. These measurements revealed a transition in exoplanet bulk composition at $\approx$1.5~\Re, with smaller planets having bulk densities consistent with rock and larger planets having extended low-density envelopes \citep{Weiss14,Rogers15}.

The distribution of \Kepler planets as a function of size, orbital period, and other properties encodes key aspects of planet formation physics including the growth of solid cores, the accretion/loss of gaseous envelopes, and the extent to which planets migrate. Insight into these processes requires accurate knowledge of host star properties. Until recently, the properties of the vast majority of \Kepler planet host stars were based on photometry alone, i.e. from the \Kepler Input Catalog (KIC) and its updates \citep{Brown11,Huber14,Mathur17}. Importantly, stellar radii \Rstar determined from photometry are uncertain at the $\approx$40\% level, which hides important features in the planet population. 

To address the limitations of photometric stellar properties, our team conducted the California-\Kepler Survey (CKS), which obtained high-resolution optical spectra of 1305 planet hosting stars \citep[][P17 hereafter]{Petigura17}. Among other properties, these spectra enabled more precise stellar radii with $\approx$11\% precision \citep[][J17 hereafter]{Johnson17}. In \cite{Fulton17}, F17 hereafter, we recomputed planet occurrence given these improved properties and found that the radius distribution of small planets is bimodal with a paucity of planets between 1.5 and 2.0~\Re. In previous studies, this feature was washed out to large \Rp uncertainties. The radius gap occurs at the transition radius separating planets with and without gaseous envelopes.

A gap in the radius distribution was predicted by several groups who considered the effect of photo-evaporation on planetary envelopes by X-ray and extreme ultraviolet (XUV) radiation (\citealt{Lopez13b,Owen13,Jin14,Chen16}). The observation of the radius gap lends much credibility to photoevaporation as a key process that sculpts the population of sub-Jovian class planets. However, while photoevaporation is a leading theory explaining this feature, alternative mechanisms have been proposed, such as mass loss powered by luminosity of a planet's cooling core \citep{Ginzburg18}.

The apparent width of the radius gap in F17 was $\approx$25\% in \Rp. Because the gap was only marginally wider than the \Rp uncertainties ($\approx$13\%), its true width and depth was uncertain. Indeed, \cite{VanEylen17} studied a smaller sample of $\sim$100 planets with percent-level \Rp precision enabled by asteroseismology, and suggested that the gap may be wider an deeper than it appears in F17.

Here, we re-examine the planet population at higher resolution by incorporating recently released parallax measurements from ESA's \Gaia mission \citep{Gaia16a}. Launched in 2013, \Gaia is conducting an all-sky astrometric survey of $\sim10^9$ stars. {\em Gaia's} first data release (DR1) included 14 months of \Gaia measurements, leveraging the Tycho catalog to constrain proper motions \citep{Gaia16b,Lindegren16}. \Gaia DR1 included parallax measurements of only a handful of \Kepler planet hosts and were not precise enough to improve radii over those from spectroscopy alone. \Gaia DR2 \citep{Gaia18}%
\footnote{Released on 2018-04-25}
is the first \Gaia-only catalog and is based on 22 months of observations. DR2 provides sub-1\% distances to the majority of \Kepler planet hosts, enabling more precise stellar and planetary radii.

Our paper is organized as follows: Section~\ref{sec:initial-sample} describes our sample selection. We derive new stellar radii in Section~\ref{sec:srad}, with \stat{srad-ferr-med}\% precision. In Section~\ref{sec:planet-population}, we derive new planet radii and examine the exoplanet population with our high-resolution sample. We also characterize astrophysical spread in the planet size distribution and note correlations between the exoplanet population and stellar mass. We conclude in Section~\ref{sec:conclusions}, connecting our observations to planet formation theory.

\section{Initial Sample Selection}
\label{sec:initial-sample}
We began with the sample of planet host stars in the CKS sample. The CKS sample selection, spectroscopic observations, and spectroscopic analysis are described in detail in P17. In brief, the sample was initially constructed by selecting all \Kepler Objects of Interest (KOIs) brighter than \Kp = 14.2~mag. A KOI is a \Kepler target star which showed periodic photometric dimmings indicative of planet transits. However, not all KOIs have received the necessary follow-up attention needed to confirm the planets. Over the course of the CKS project, we included additional targets to cover different planet populations, including multi-candidate hosts, ultra-short period candidates, and habitable zone candidates. 

We cross-matched the CKS sample with the \Gaia DR2 catalog by querying all \Gaia sources within 1~arcsec of the KIC coordinates. In rare cases, \Gaia detected more than one source within 1~arcsec, and we selected the source with the smallest difference between $G$ and $\Kp$ magnitudes. We cross-matched \stat{cks-gaia-star-xmatch-count} targets in this way.

\section{Stellar Radii}
\label{sec:srad}
\subsection{Introduction}
\label{sec:srad-intro}
Our re-derived stellar radii (\Rstar) follow from the Stefan--Boltzmann law,
\begin{equation}
\label{eqn:rstar}
\Rstar = \left(\frac{\Lbol}{4\pi\sigmasb\teff^4}\right)^{1/2},
\end{equation}
where \Lbol is the bolometric stellar luminosity, \sigmasb is the Stefan--Boltzmann constant, and \teff is the effective temperature. \Lbol is related to bolometric magnitude $\Mbol$ via
\begin{equation}
\label{eqn:lbol}
\Lbol = L_0 10^{-0.4 \Mbol},
\end{equation}
where $L_0$ is defined to be $L_0 = 3.0128 \times 10^{28}$~W.%
\footnote{See IAU 2015 Resolution B2 and \cite{Mamajek15}.}
\Mbol may be measured from a single broadband photometric apparent magnitude $m$, if the distance modulus $\mu$, line-of-sight extinction $A$, and bolometric correction $BC$ are known:
\begin{equation}
\label{eqn:mbol}
\Mbol = m - A - \mu + BC.
\end{equation}
Therefore, our derived stellar radii depend on five parameters: $m$, $A$, $\mu$, \teff, $BC$. We discuss the provenance of these in parameters in Sections~\ref{sec:photometry}--\ref{sec:bolometric} along with their respective contributions to the \Rstar error budget, which are summarized in Table~\ref{tab:error-budget}. Section~\ref{sec:detailed-modeling} explains our detailed modeling of \Rstar, which closely follows that of \cite{Huber17}. We validate our stellar radii through a comparison with asteroseismology  in  Section~\ref{sec:astero}. We also compare our radii to the purely spectroscopic measurements of J17 in Section~\ref{sec:comparison-cks2} and to those computed by the \Gaia project in Section~\ref{sec:comparison-gaia2}.

\subsection{Photometry}
\label{sec:photometry}
We used $K$-band photometric measurements because dust extinction is less severe in the infrared (see Section~\ref{sec:extinction}). The Two Micron All-Sky Survey (2MASS; \citealt{Skrutskie06}) measured $m_K$ for our target stars with a median precision of \stat{mk-err-med}~mag, which corresponds to $\approx$1\% errors in \Rstar.

We elected to use a single photometric band so that our \teff constraints would depend only on spectroscopy. Compared broadband colors, spectroscopy has the advantage that it yields more precise temperatures that are insensitive to reddening. For \Kepler field stars, there are significant degeneracies between reddening and photometric \teff that result in uncertainties of $\approx$200~K (see \citealt{Pinsonneault12} and P17). Given that \teff uncertainties often dominate the final \Rstar uncertainty, we restricted our analysis to $m_K$.

As an aside, we expect that \Gaia DR2 will transform our understanding of the 3D distribution of dust in the Milky Way Galaxy. This will reduce reddening-\teff degeneracies for \Kepler field stars, and result in improved measurements of \teff from broadband colors. However, such a dust modeling effort is beyond the scope of this work. 

\subsection{Extinction}
\label{sec:extinction}

We consulted the 3D dimensional dust map of \cite{Green18} to quantify and correct for $K$-band extinction. The map tabulates reddening in PS1 and 2MASS passbands as a function of a function of galactic latitude, galactic longitude, and distance. Our median target has a $E(B-V)$ reddening of $\stat{ebv-med}$~mag.

To convert between between $E(B-V)$ and $A_\lambda$, one must multiply $E(B-V)$ by an extinction vector $R_\lambda$. \cite{Green18} adopted $R_\lambda$ from \cite{Schlafly16} who studied the variation in observed stellar colors with reddening.  Unfortunately, the \cite{Schlafly16} methodology is insensitive to the gray component of the extinction curve, i.e. $R_\lambda \rightarrow R_\lambda + b$. As a matter of convenience, \cite{Green18} resolved this ambiguity by setting $R_{W2}$ = 0, which implies $R_K$ = 0.161. However, if one adopts $A_H / A_K$, one may derive $b$ by solving the following system of equations:

\begin{eqnarray}
A_H & = & (R_H + b) E(B-V) \nonumber \\
A_K & = & (R_K + b) E(B-V) \nonumber.
\end{eqnarray}
We adopted $A_H / A_K$ = 1.74 from \cite{Nishiyama08}, which yields $b$ = 0.063 and $R_K$ = 0.224. The value of $A_H / A_K$ itself is uncertain. As a point of reference, \cite{Indebetouw05} found $A_H / A_K$ = 1.55, which yields $b$ = 0.141 and $R_K$ = 0.302. To account for the uncertainty in $R_K$, we add 30\% additional fractional uncertainty to $A_K$.

The expected $K$-band extinction ranges from $A_K$ = \stat{ak-min}--\stat{ak-max}~mag, with a median value of $A_K$ = \stat{ak-med}~mag. The low typical extinction highlights the advantage of $K$-band. Neglecting extinction entirely would result in a $\approx$0.5\% error in \Rstar for our median target, which is smaller than other terms in the \Rstar error budget. For completeness, we incorporated $A_K$ derived from the \cite{Green18} maps into our radius calculations (Section~\ref{sec:detailed-modeling}).

\subsection{Parallaxes}
\label{sec:parallax}
We used parallaxes from \Gaia DR2 and required that the parallax uncertainties be smaller than 10\%. The median parallax precision of the remaining \stat{cks-gaia-star-count} stars is \stat{cks-gaia-sparallax-ferr-med}~\% and contributes \stat{cks-gaia-sparallax-ferr-med}~\% to our \Rstar error budget. The \Gaia team recommends adopting systematic error floor of 0.1~mas, which accounts for zero-point and spatially correlated systematics. Fortunately, the \Kepler field is one of the best characterized regions of the sky and independent methods may be used to measure and correct for these systematics. \cite{Zinn18} used precise distances to asymptotic giant branch (AGB) stars to measure offsets in \Gaia parallaxes for the \Kepler field and found that the \Gaia parallaxes were too small by 0.053~mas. In this work, we apply a correction of +0.053~mas to the \Gaia parallaxes to account for this offset.

\subsection{Effective Temperatures}
\label{sec:teff}
Stellar effective temperatures factor into our measurement of stellar radii in two ways: through the Stefan--Boltzmann law (Section~\ref{sec:srad-intro}) and through the bolometric corrections (Section~\ref{sec:bolometric}). We used the CKS spectroscopic \teff which have an internal precision 60~K (P17). We note that offsets of $\approx$100~K are often observed when comparing different spectroscopic catalogs as well as when comparing spectroscopic temperatures to temperatures determined by other techniques, such as the infrared flux method or interferometry \citep{Brewer16}. Therefore, these temperatures are accurate on an absolute scale to $\approx$100~K. However, since our radii are all derived using CKS \teff, the \teff precision, rather than its absolute accuracy, factors into the precision of our stellar radii. A precision of 60~K corresponds to $\approx2\%$ errors on \Rstar.

\subsection{Bolometric Corrections}
\label{sec:bolometric}
With $m_k$, $A_K$, and $\mu$ we may compute absolute $K$-band magnitude, $M_K$. Converting $M_K$ to $\Mbol$ requires a bolometric correction $BC_K$. We computed $BC_K$ using the isoclassify package \citep{Huber17} which interpolates over the MIST grid of bolometric corrections \citep{Choi16}. For each star, we found the range of $BC_K$ consistent with our spectroscopically determined \teff and \fe. The uncertainties on \teff dominate the uncertainty of $BC_K$ because \teff has the largest influence on shape of the stellar SED. For a Sun-like star a  60~K uncertainty translates to a $\approx$0.03~mag error on $BC_K$ or $\approx$1.5\% errors in \Rstar. Errors on $BC_K$ stemming from uncertain \logg and \fe are negligible by comparison.

We note that \teff errors enter into the Stefan-Boltzmann Law and $BC_K$ in ways that largely cancel. For our stars, $K$-band probes the Rayleigh-Jeans tail of the SED, where flux scales like \teff. Therefore, at a fixed $m_K$, $\Lbol \propto \teff^{3}$, which is largely canceled by the $\teff^{-4}$ term in Equation~\ref{eqn:rstar}.

The bolometric corrections also include model-dependent errors. \cite{Huber17} assessed these errors by comparing stellar radii derived from the MIST grids to those derived using the BASTA grids \citep{Casagrande14} and estimated 0.03~mag errors. As with \teff, we expect these model-dependent errors to be largely common-mode and are thus more relevant for the accuracy rather than the precision of our stellar radii.

\subsection{Detailed Modeling}
\label{sec:detailed-modeling}
\begin{figure*}
\centering
\includegraphics[width=0.45\textwidth]{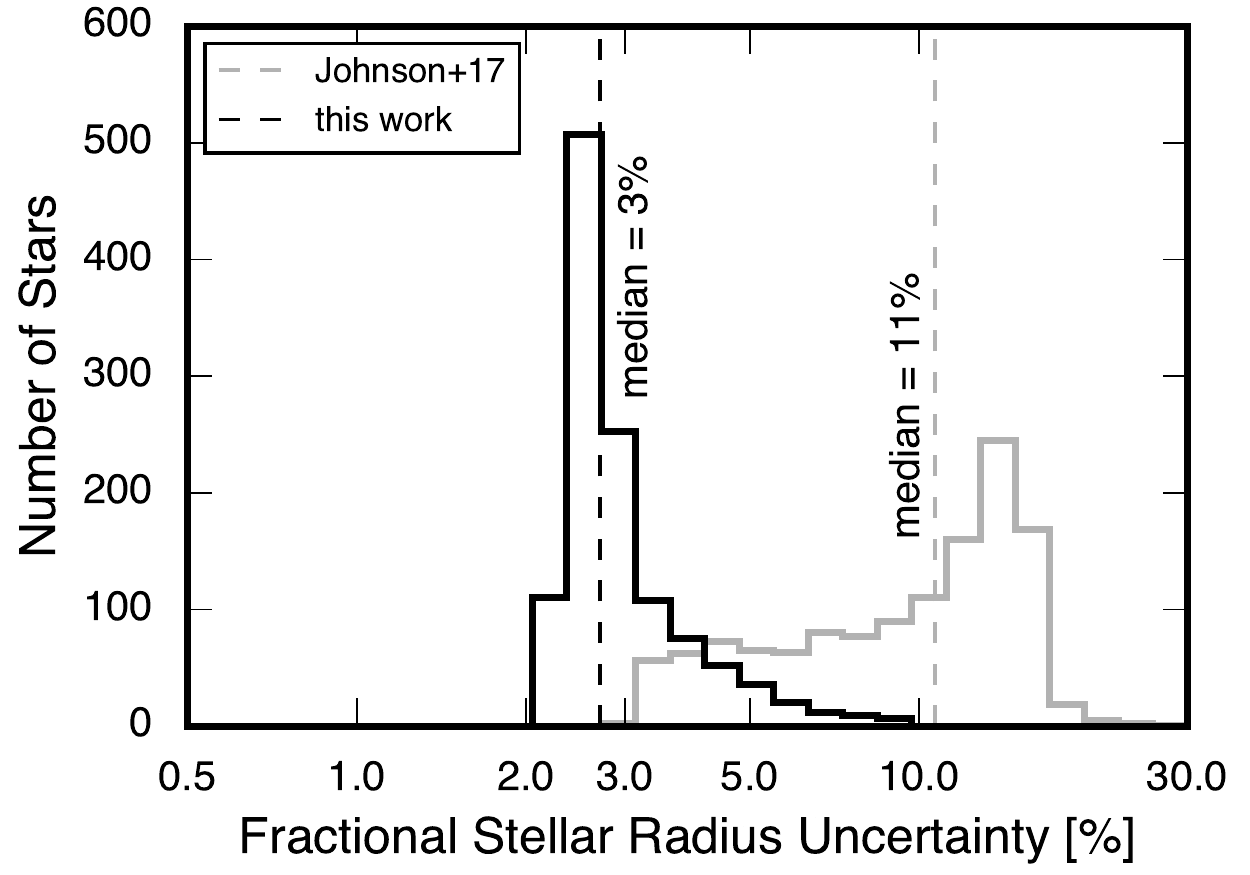}
\includegraphics[width=0.45\textwidth]{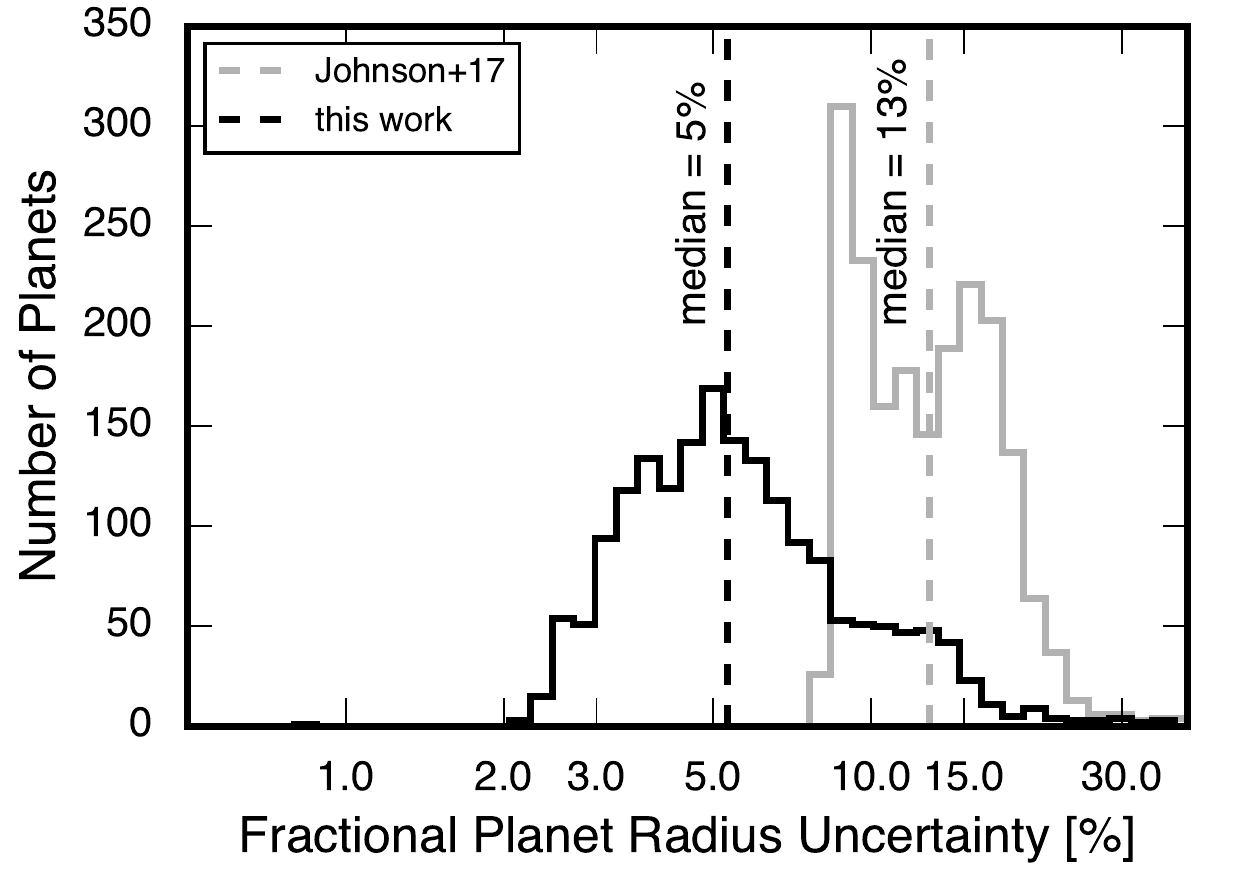}
\caption{{\em Left:} Distribution of fractional stellar radius uncertainties from this work (black) compared to those from \citet{Johnson17} (grey). {\em Right:} Same as {\em left} but comparing fractional planet radius uncertainties.}
\label{fig:err-hist}
\end{figure*}

In the previous sections, we enumerated the various measurements that we used to compute \Rstar and estimated their final contribution to the \Rstar error budget, which we summarize in Table~\ref{tab:error-budget}. To compute the radii, we used the isoclassify package in its ``direct'' mode \citep{Huber17}. For each star, we provided isoclassify with \teff, \fe, $\pi$, and $m_K$. Then, isoclassify computed the posterior probability on \Rstar using Equations~\ref{eqn:rstar}--\ref{eqn:mbol}. As an intermediate step, isoclassify must infer a distance given the parallax measurement. This is done using Bayesian inference, incorporating an exponentially decreasing volume density prior with a length scale of 1.35 kpc, as recommended by \cite{Astraatmadja16}. Figure~\ref{fig:err-hist} shows the distribution of the formal \Rstar precision, which have a median value of \stat{srad-ferr-med}\%. The radii are provided in Table~\ref{tab:star}. 

One advantage of deriving radii from the Stefan-Boltzmann law is they are minimally model-dependent; they rely on models only for the bolometric corrections (Section~\ref{sec:bolometric}). A disadvantage is that this analysis does not constrain stellar mass and age. We performed a parallel analysis with isoclassify using its ``grid'' mode. In this mode, isoclassify computes the range of masses, radii, and ages that are consistent with the observational constraints and the MIST isochrone grids. We include these parameters in Table~\ref{tab:star} as a matter of convenience. We caution that their formal uncertainties do not include systematic uncertainties associated with the MIST models. Such uncertainties are likely largest for the coolest stars in our sample. 

Finally, we must consider the effects of flux contamination from unresolved binaries on our radius measurements. If a given target star has a companion within the 2MASS software aperture, typically 4 arcsec \citep{Skrutskie06}, the target star appears brighter and we infer a larger radius. In Section~\ref{sec:occurrence}, we screen out contaminating sources using the \Gaia source catalog, which has an effective angular resolution of 0.4~arcsec \citep{Arenou18}, and existing high resolution follow-up imaging. As an additional check, we ran isoclassify in the ``grid'' mode while providing \teff, \logg, \fe, and  $m_K$ constraints, but no parallax constraints. For each star, isoclassify returned a parallax consistent with the input constraints and the MIST models. If the ``isochrone parallax'' is significantly larger than the \Gaia parallax the star is likely an unresolved binary. We include this ``isochrone parallax'' in Table~\ref{tab:star} and recommend using radii where the two parallax measurements are consistent to four sigma.

\subsection{Validation with Asteroseismic Radii}
\label{sec:astero}
\begin{figure*}
\centering
\includegraphics[width=0.45\textwidth]{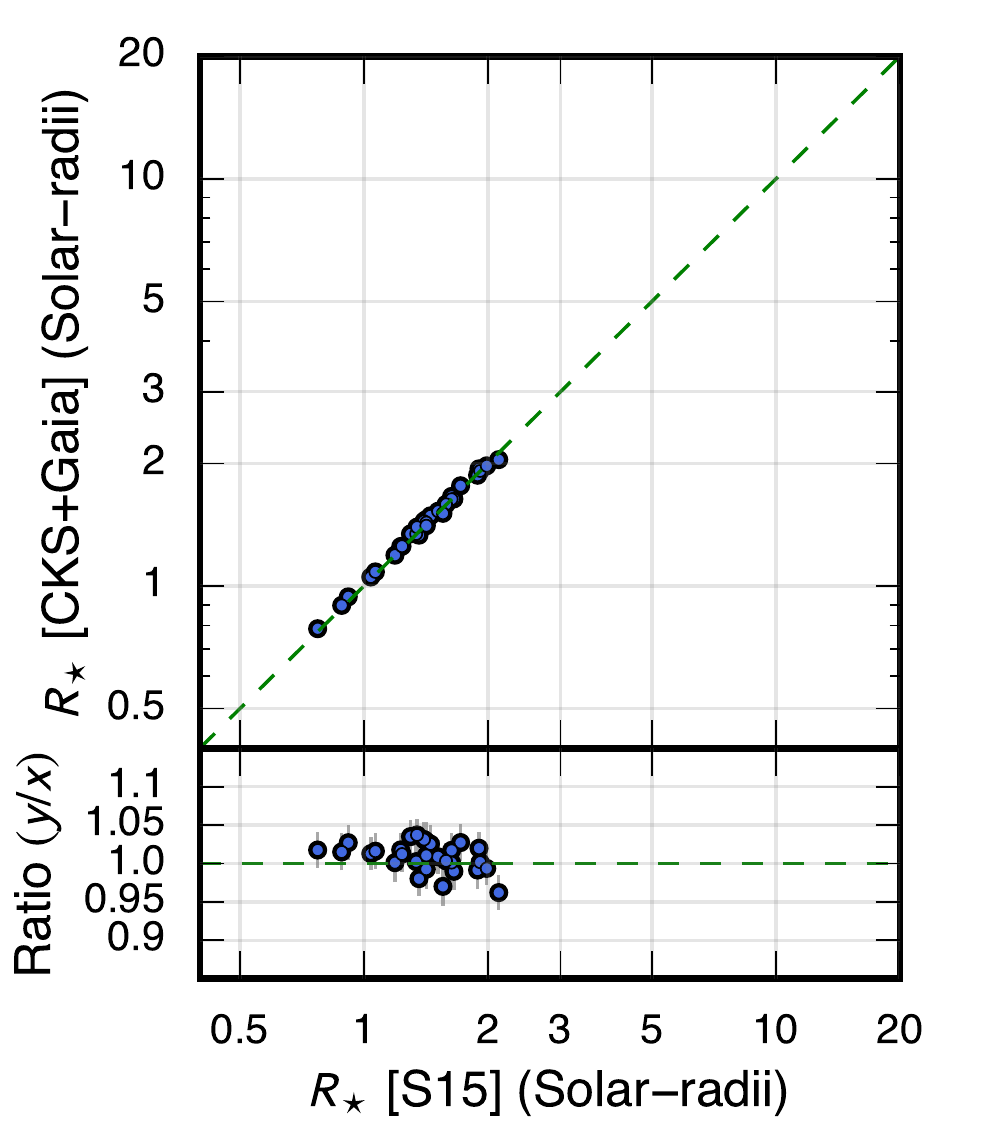}
\hspace{-0.25in}
\includegraphics[width=0.45\textwidth]{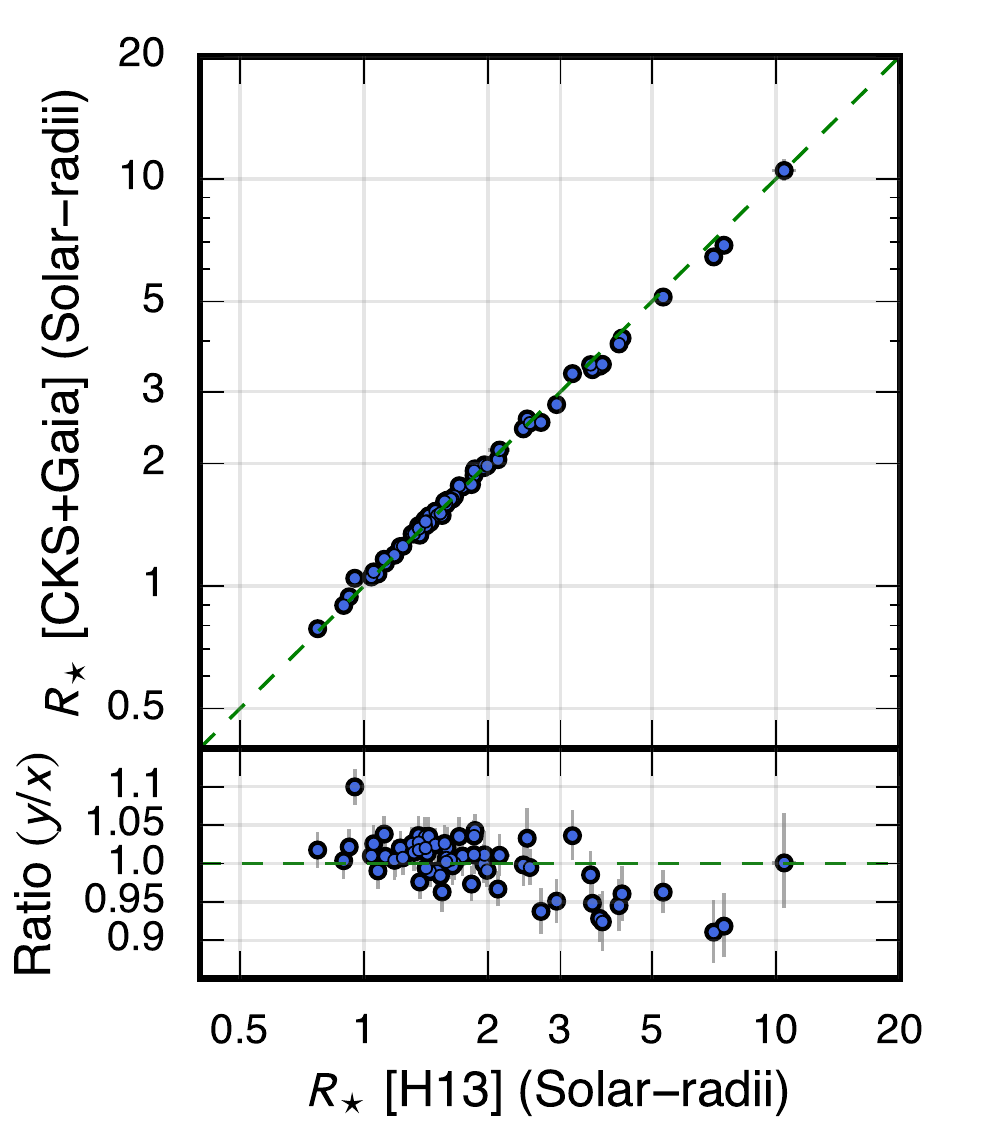}
\caption{{\em Left:} Comparison of stellar radii derived from asteroseismology (\citealt{Silva15}; S15) and spectroscopy+astrometry (this work) for \stat{cks-s15-count} stars in common. Equality is represented by the dashed green line. Our radii are \stat{cks-s15-srad-mean} on average and there is a \stat{cks-s15-srad-std} RMS dispersion in the ratios. {\em Right:} same as {\em left} but comparing our radii to \citealt{Huber13a} (H13). Our radii are \stat{cks-h13-srad-mean} on average and there is a \stat{cks-h13-srad-std} RMS dispersion in the ratios. \label{fig:srad-s15-h13}}
\end{figure*}

As in Paper-II, we assessed the final precision and accuracy of our stellar radii with a comparison to stellar radii derived using asteroseismology. We first compared against radii from \cite{Silva15}, S15 hereafter, who performed an asteroseismic analysis of 33 \Kepler planet hosts and achieved a median radius uncertainties of $\approx$1\%. Importantly, the S15 analysis modeled individual oscillation frequencies, which achieves higher precision than simpler asteroseismic scaling relationships. We compared the stellar radii for the \stat{cks-s15-count} CKS stars in common with the S15 study (Figure~\ref{fig:srad-s15-h13}). Our radii are \stat{cks-s15-srad-mean} on average, with a \stat{cks-s15-srad-std} RMS scatter in the ratio, which is consistent with the quadrature sum of the formal uncertainties of both sets of radii. 

Because the S15 radii span a narrow range in \Rstar of  0.7--2.0~\Rsun, we performed a second comparison against radii from \cite{Huber13a}, H13 hereafter, which span 0.7--10~\Rsun. H13 relied on scaling relationships using the small frequency separation $\delta \nu$ and peak frequency $\nu_{max}$. These relationships are lower precision than the more detailed analysis of S15 at 3\% fractional precision. Our radii are \stat{cks-h13-srad-mean} on average, with a \stat{cks-h13-srad-std} RMS scatter in the ratio, which is consistent with the quadrature sum of the formal uncertainties of both sets of radii. 

Our comparisons with S15 and H13 show that our stellar radius precision is comparable to, or smaller than, those from asteroseismology. In principle our methodology for measure stellar radii can be used to test systematics in the asteroseismic scaling relationships as in \citep{Huber17}. Such an effort is beyond the scope of this work.

\subsection{Comparison with \cite{Johnson17} Radii}
\label{sec:comparison-cks2}
Figure~\ref{fig:srad-j17-gaia2} compares our radii against those from J17, which relied on spectroscopy alone. The RMS scatter in the ratios is \stat{cks-j17-srad-std}, which is consistent with the 11\% median uncertainty quoted in J17. We also note that the J17 radii on average fall below the one-to-one line between 1 and 3~\Rsun. We observed this trend in J17 when comparing the J17 radii to asteroseismic radii. It is not surprising that we observe this same trend in a larger sample given our new radii closely track asteroseismology. This demonstrates the potential for \Gaia to serve as a benchmark with which to test synthetic spectra and model atmospheres. We also note a handful of outliers in the comparison. These could be due to stars with unresolved companions contributing extra $K$-band flux and making the CKS+Gaia radii seem larger. They may also be due to rare and unknown failure modes in the spectroscopic analysis of P17. 

\subsection{Comparison with Gaia DR2 Stellar Radii}
\label{sec:comparison-gaia2}
\begin{figure*}
\centering
\includegraphics[width=0.45\textwidth]{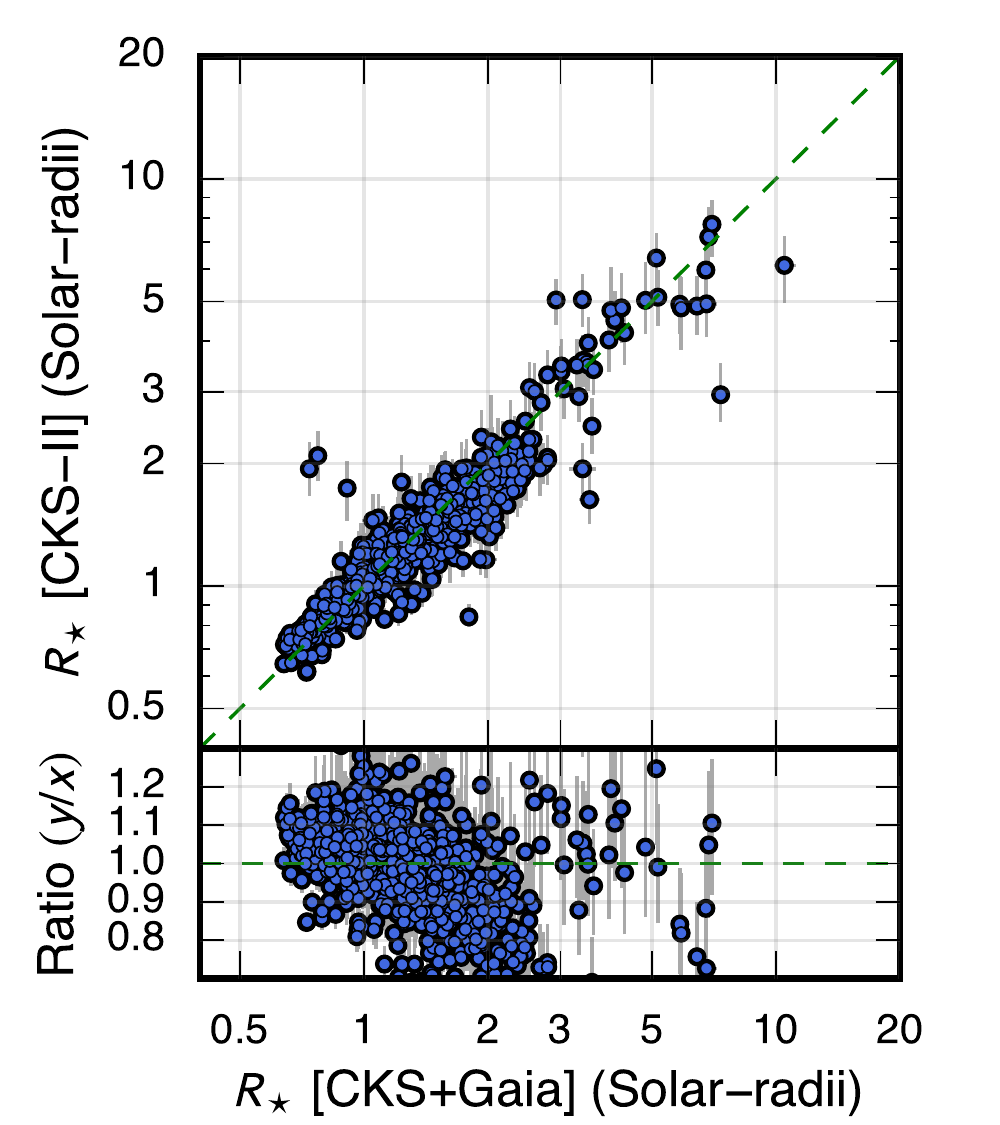}
\includegraphics[width=0.45\textwidth]{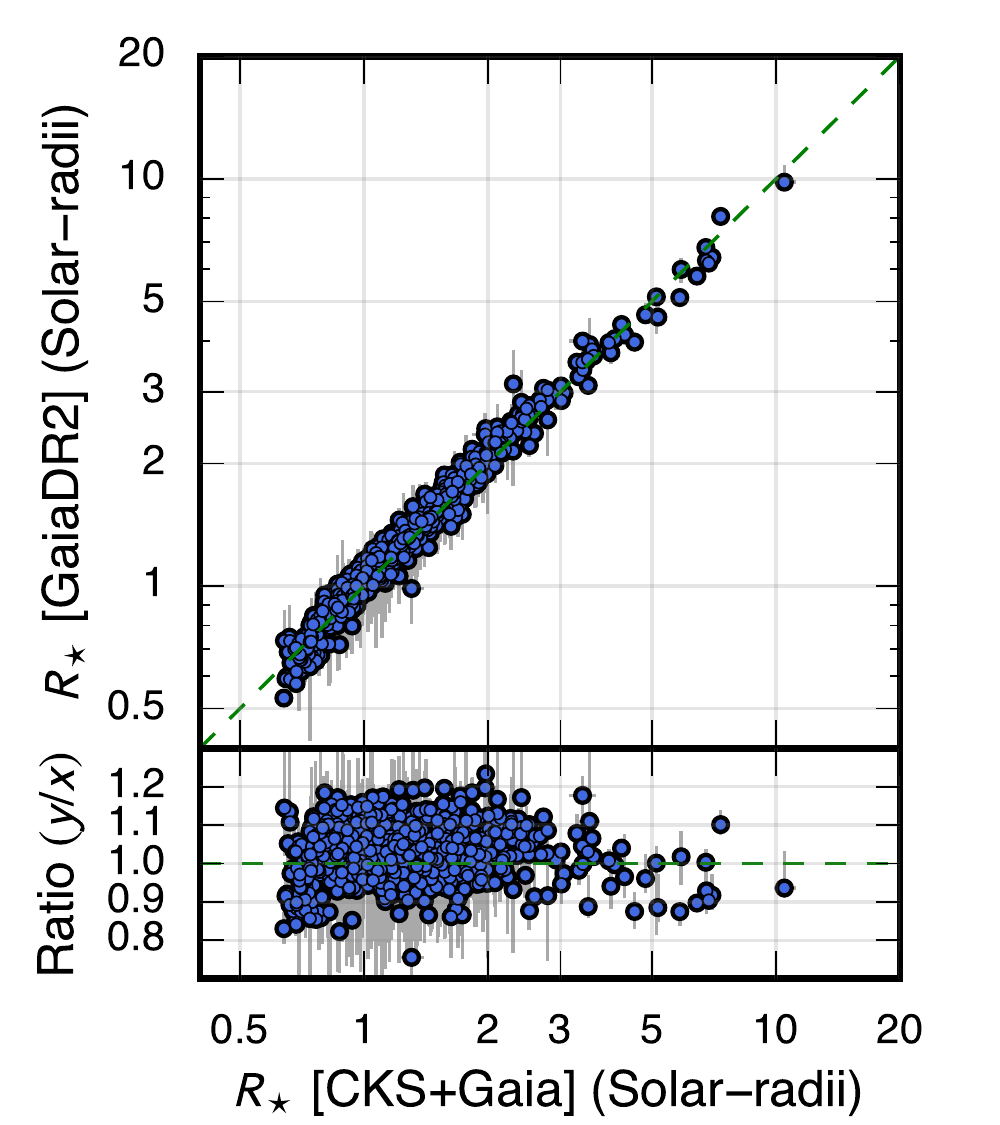}
\caption{{\em Left:} Comparison of stellar radii derived from spectroscopy+astrometry (this work) and spectroscopy alone \citep[][CKS-II]{Johnson17}. The CKS-II radii are \stat{cks-j17-srad-mean} on average and there is a \stat{cks-j17-srad-std} RMS dispersion in the ratios. {\em Right:} same as {\em left} but comparing our radii to the \Gaia DR2 radii. The \Gaia DR2 radii are \stat{cks-gaia2-srad-mean} on average and there is a \stat{cks-gaia2-srad-std} RMS dispersion in the ratios. \label{fig:srad-j17-gaia2}}
\end{figure*}

The \Gaia project also provided radii based on SED modeling that fits for effective temperature, extinction, and radius given the known distance. Figure~\ref{fig:srad-j17-gaia2} compares our radii with the \Gaia project radii for \stat{cks-gaia2-count} stars in common. On average \Gaia DR2 radii are \stat{cks-gaia2-srad-mean}  than ours with a \stat{cks-gaia2-srad-std} RMS scatter in the ratio, which is consistent with the formal median uncertainty of \stat{cks-gaia2-srad-err-mean} reported in \Gaia DR2.

\begin{deluxetable}{lrr}
\tablecaption{Star and Planet Properties\label{tab:error-budget}}
\tablewidth{0pt}
\tablehead{
  \colhead{Parameter}   & 
  \colhead{Median Value} &
  \colhead{Median Uncertainty}
}
\startdata
\teff         & \stat{steff-med}~K                & 60~K                       \\
$m_K$         & \stat{mk-med}~mag                 & 0.02~mag                   \\
$A_K$         & \stat{ak-med}~mag                 & 0.004~mag                  \\
$\pi$         & \stat{cks-gaia-sparallax-med}~mas & \stat{cks-gaia-sparallax-ferr-med}\% \\ 
$\mu$         & \stat{cks-gaia-distmod-med}~mag   & \stat{cks-gaia-distmod-err-med}~mag \\ 
$BC$          & $-$1.46~mag                         & 0.03~mag                   \\
$\Rstar$      &  \stat{srad-med}~\Rsun            & \stat{srad-ferr-med}\%     \\ 
$\Rp/\Rstar$  & \stat{ror-med}\%                  & \stat{ror-ferr-med}\%      \\
$\Rp$         & \stat{prad-med}~\Re               & \stat{prad-ferr-med}\%     \\
\enddata
\end{deluxetable}

\section{Planet Population}
\label{sec:planet-population}

\subsection{Distribution of Detected Planets}
\begin{figure*}[ht]
\centering
\includegraphics[width=0.49\textwidth]{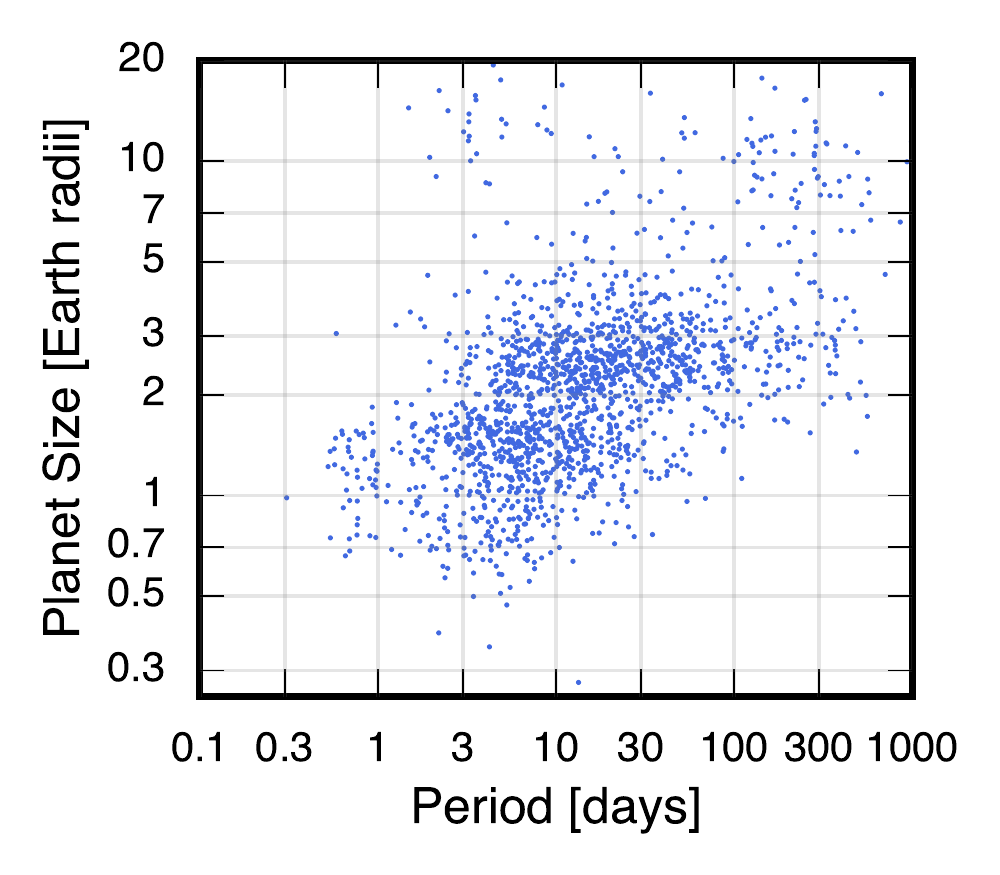}
\includegraphics[width=0.49\textwidth]{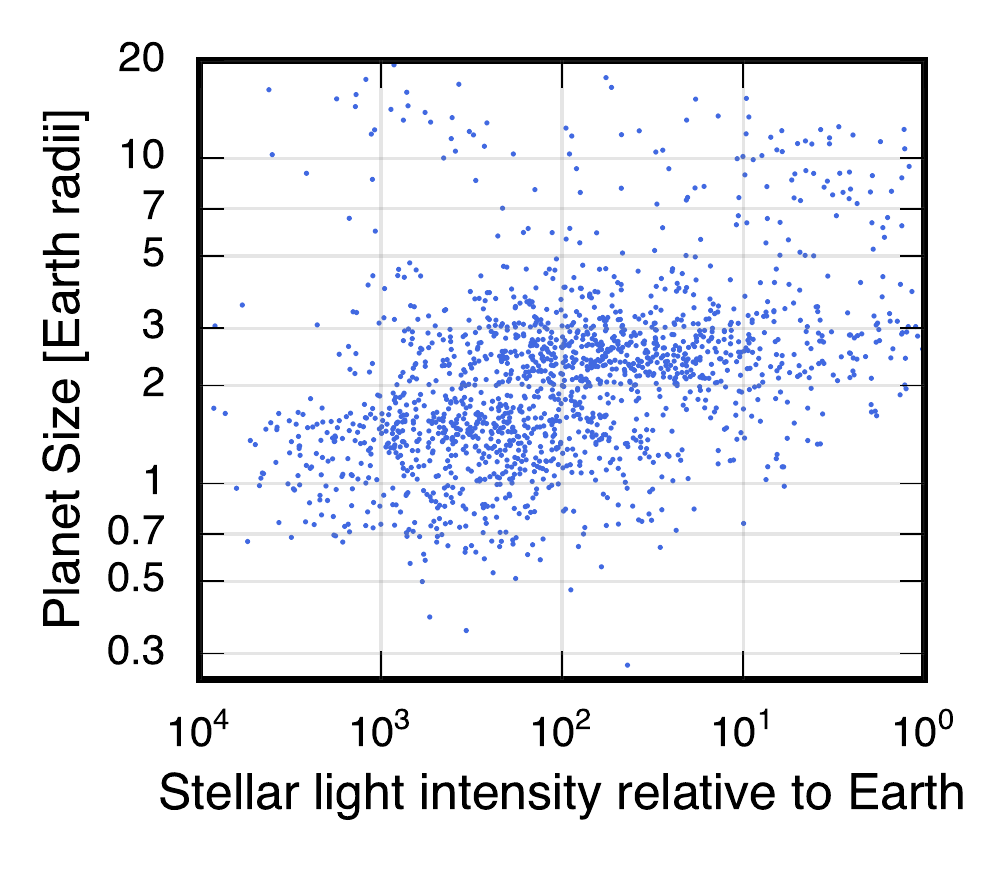}
\caption{{\em Left:} distribution of planet radii and orbital periods. {\em Right:} same as {\em left} but with insolation flux relative to Earth on the horizontal axis. In both plots, an underdensity of points appears between 1.5 and 2.0~\Re.\label{fig:per-insol}}
\end{figure*}

Using our updated stellar radii we derived planet radii using the values of \Rp/\Rstar tabulated in \cite{Mullally15}. We also computed the incident stellar flux \Sinc using our updated \Rstar and \teff.  These \Rp and \Sinc measurements are listed in Table~\ref{tab:planet}. Figure~\ref{fig:per-insol} shows the distribution of planets in the $P$-\Rp and \Sinc-\Rp planes.

As in F17, we observe a narrow gap separating two populations of planets at $\approx$2 \Re. While the gap is clearly visible in this sample of \stat{planet-count} planets, we caution that the distribution of detected planets does not convey the underlying distribution of planets, due to selection effects that we discuss in Section~\ref{sec:occurrence}.

\subsection{Intrinsic Distribution of Planets}
\label{sec:occurrence}

Here, we measure planet occurrence, the number of planets per star, as a function of $P$, \Rp, and \Sinc. In order to measure the intrinsic distribution of planets, we must account for selection effects in the construction of the CKS target list, geometrical transit probability, and pipeline completeness. Our methodology follows that of F17.

We first identified a subset of CKS planets drawn from a well-defined population of parent stars by applying the following cuts to our planet sample:

\begin{enumerate}
\item {\em Stellar brightness.} We restricted our sample to the magnitude-limited CKS subsample, where \Kp $<$ 14.2 mag.

\item {\em Stellar radius.} We restricted our analysis to dwarf stars where

\begin{equation}
\log_{10} \left( \frac{\Rstar}{\Rsun} \right) < \left(\frac{\teff - 5500\, \K}{4000\, \K} \right) + 0.2.
\end{equation}

\item {\em Stellar effective temperature.} We restricted our planet sample to stars with \teff = 4700--6500~K, where the CKS temperatures are reliable.

 \item {\em Isochrone parallax.} For each star, we computed an ``isochrone parallax'' based on \teff, \logg, \fe, and $m_K$ (see Section~\ref{sec:detailed-modeling}). We removed stars where the \Gaia and isochrone parallaxes differed by more than 4$\sigma$, due to likely flux contamination by unresolved binaries.

\item {\em Stellar dilution (Gaia).} Dilution from nearby stars can also alter the apparent planetary radii. For each target, we queried all \Gaia sources within 8~arcsec (2 \Kepler pixels) and computed the sum of their $G$-band fluxes. The ratio between this cumulative flux and the target flux $r_8$ approximates the $\Kp$-band dilution for each transiting planet. We required that $r_8 < 1.1$.

\item {\em Stellar dilution (imaging).} \cite{Furlan17} compiled high-resolution imaging observations performed by several groups. When a nearby star is detected, \cite{Furlan17} computed a radius correction factor (RCF), which accounts for dilution assuming the planet transits the brightest star. We do not apply this correction factor, but conservatively exclude KOIs where the RCF exceeds 5\%.

\item {\em Planet false positive designation.} We excluded candidates that are identified as false positives according to P17.

\item {\em Planets with grazing transits.} We excluded stars having grazing transits ($b$ $>$ 0.9), which have suspect radii due to covariances with the planet size and stellar limb-darkening during the light curve fitting.
\end{enumerate}
After applying these cuts, we are left with \stat{planet-filtered-count} planets. 

Where possible, we applied the same filters on stellar properties to the \Kepler field star population. For the stellar radius and temperature filters we used the \Gaia DR2 parameters. We could not apply the imaging cut to the parent stellar population because it relies on follow-up resources directed specifically at KOIs not at the parent parent population. After filtering, \stat{star-filtered-count} stars remain.

We calculated planet occurrence using the inverse detection efficiency methodology IDEM of F17. In brief, we account for the detection sensitivity of the survey using the injection-recovery tests performed by \citet{Christiansen15}. We calculated planet occurrence as the number of planets per star in discrete bins as
\begin{equation}
f_{\rm bin} = \frac{1}{N_{\star}}\sum_{i=1}^{n_{\rm pl,bin}} w_i.
\end{equation}
where $N_{\star}$ = \stat{star-filtered-count} and $w_i$ is the product of the inverse pipeline detection efficiency $p_{\rm det}$ and the inverse transit probability $p_{\rm tr}$ for each detected planet. Values of $w_i$, $p_{\rm det}$, $p_{\rm tr}$ are listed in Table \ref{tab:weights}.

Computing these weights requires knowledge of the distribution of radii and noise properties of stars in the parent stellar sample. As in F17, we used the Combined Differential Photometric Precision computed by the \Kepler project \citep{Mathur17} as our noise metric. Unlike F17, we used the \Rstar from \Gaia DR2 as opposed to photometric \Rstar to characterize the distribution of parent stellar radii. F17 found that plausible statistical and systematic errors of 40\% and 25\% respectively in the photometric radii of the parent stellar population led to errors in planet occurrence of up to a factor of two at 1.0~\Re. Our new occurrence measurements have the major advantage that there are negligible differences between the radii of the field stars and planet hosts; thus our occurrence measurements are up to twice as precise. 

The IDEM has been used in a number of previous works \citep[e.g.][]{Howard10, Howard12, Morton14, Fulton17}. While our results depend on the {\em relative} occurrence of planets as a function of host star mass, we wish to remind the reader that additional care is required when computing {\em absolute} occurrence in regions of low completeness. 

\citet{Hsu18} performed a comparison of various occurrence estimators including the IDEM used in this study. \citet{Hsu18} found that the IDEM is a biased estimator in the limit of low completeness. In brief, the bias arises because fluctuations due to photometric noise cause transits to appear larger/smaller, resulting in larger/smaller planet radii. However, near the detection limit, there is a bias toward detecting the apparently larger planets. \cite{Hsu18} recommends adopting a different estimator based on approximate bayesian computation (ABC). The ABC method has the advantage that it is less biased in regions of low completeness, but requires significantly more computational effort compared to the IDEM.

\begin{figure}[h]
\centering
\includegraphics[width=0.49\textwidth]{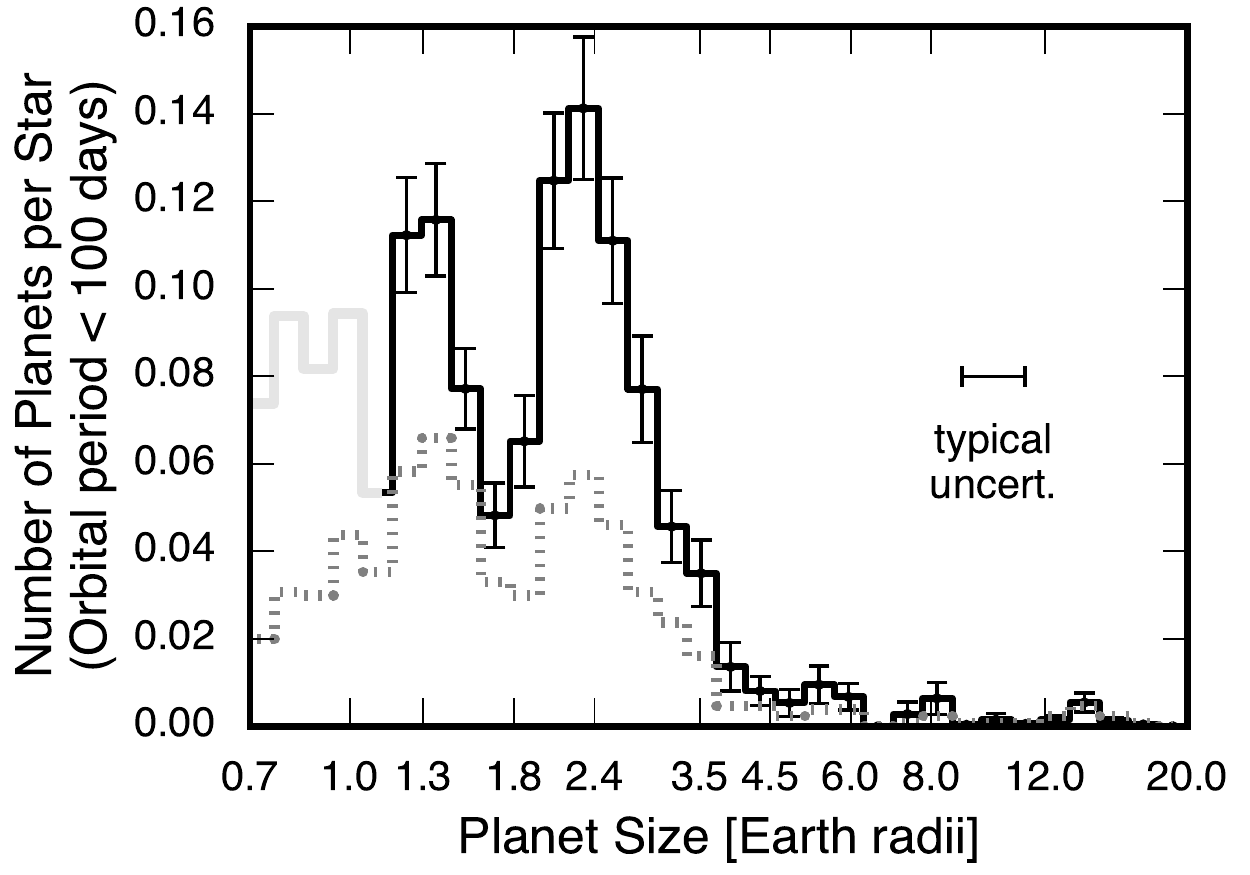}
\includegraphics[width=0.49\textwidth]{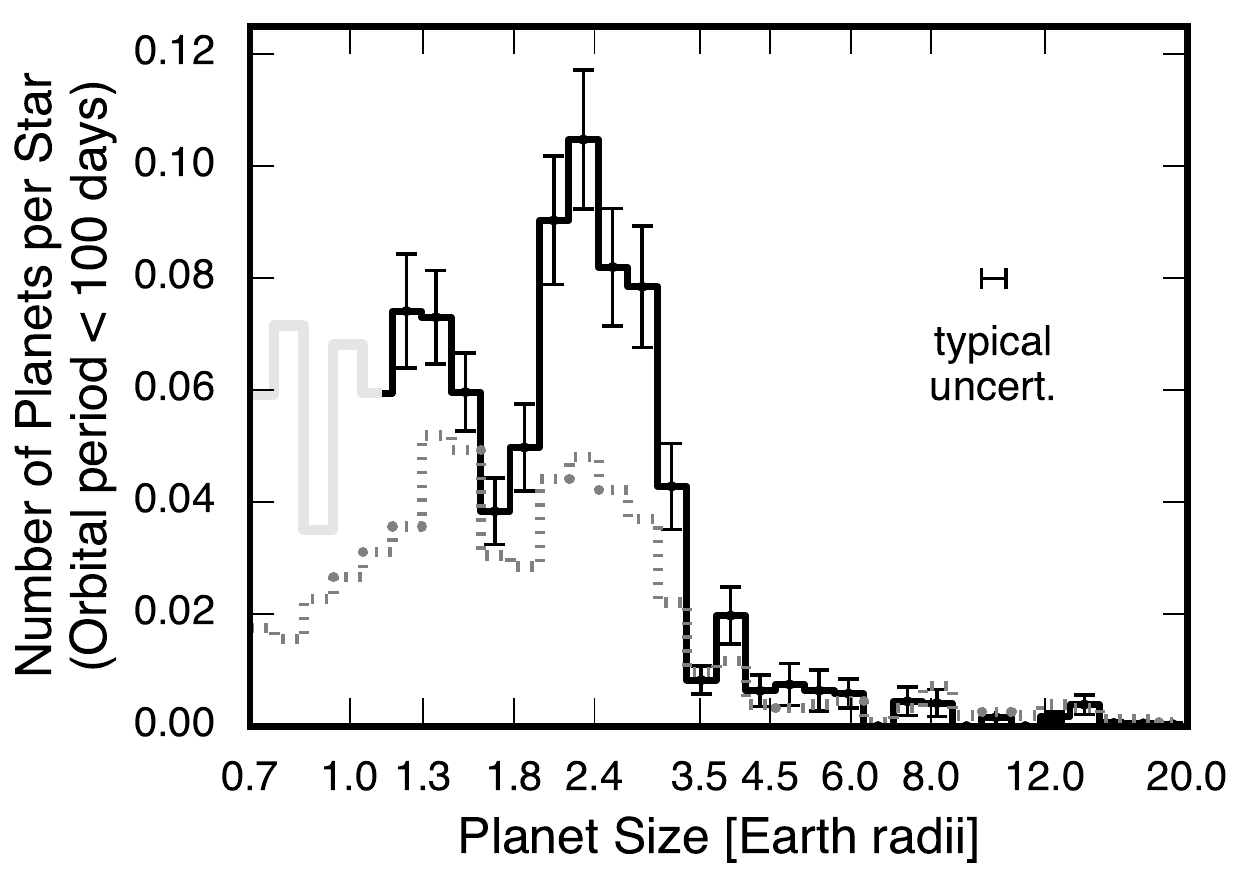}
\caption{The distribution of close-in planet sizes. The top panel shows the distribution from \citet{Fulton17} and the bottom panel is the updated distribution from this work. The solid line shows the number of planets per star with orbital periods less than 100~days as a function of planet size. A deep trough in the radius distribution separates two populations of planets with \Rp $>$ 1.7 \Re and \Rp $<$ 1.7 \Re. As a point of reference, the dotted line shows the size distribution of detected planets, before completeness corrections are made arbitrarily scaled for visual comparison. \label{fig:radius-hist}}
\end{figure}

\begin{figure*}[ht]
\centering
\includegraphics[width=0.49\textwidth]{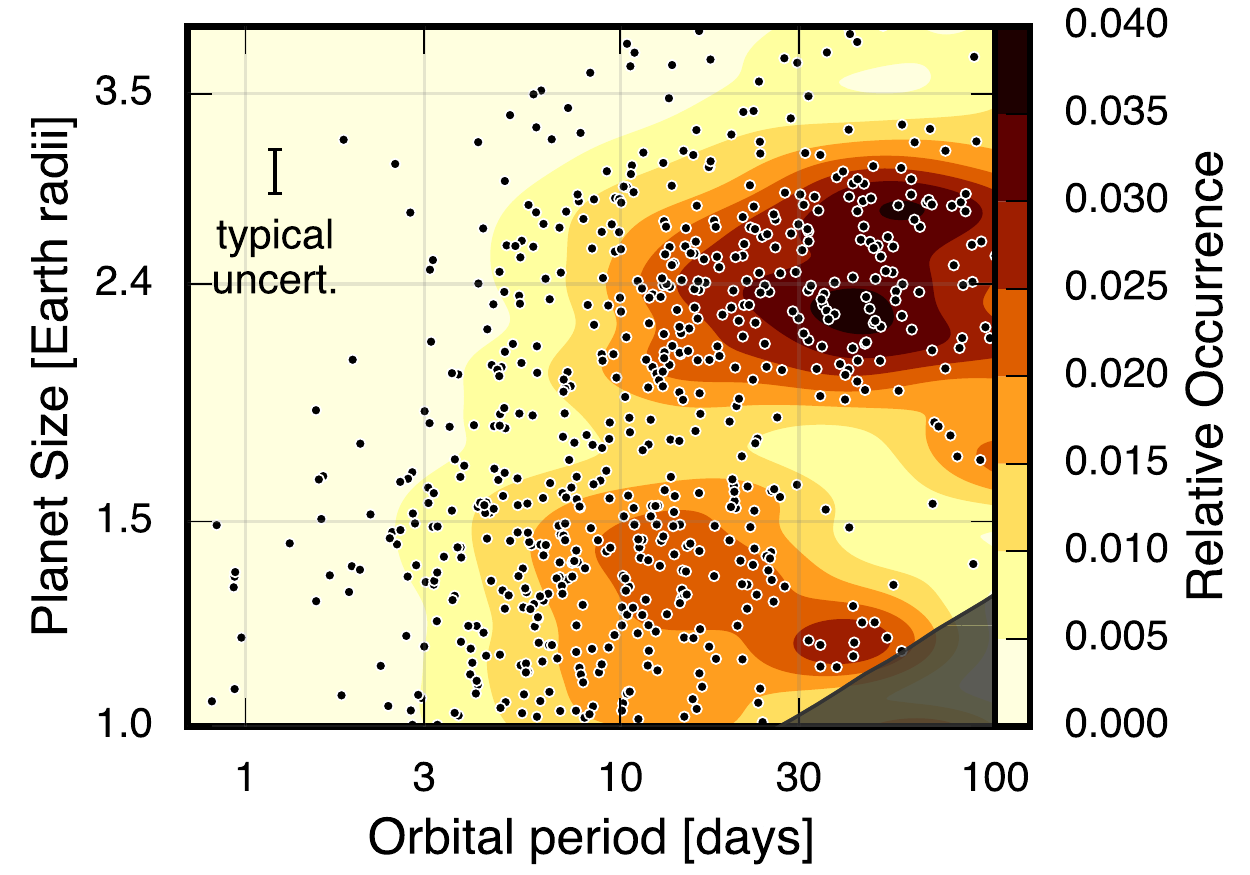}
\includegraphics[width=0.49\textwidth]{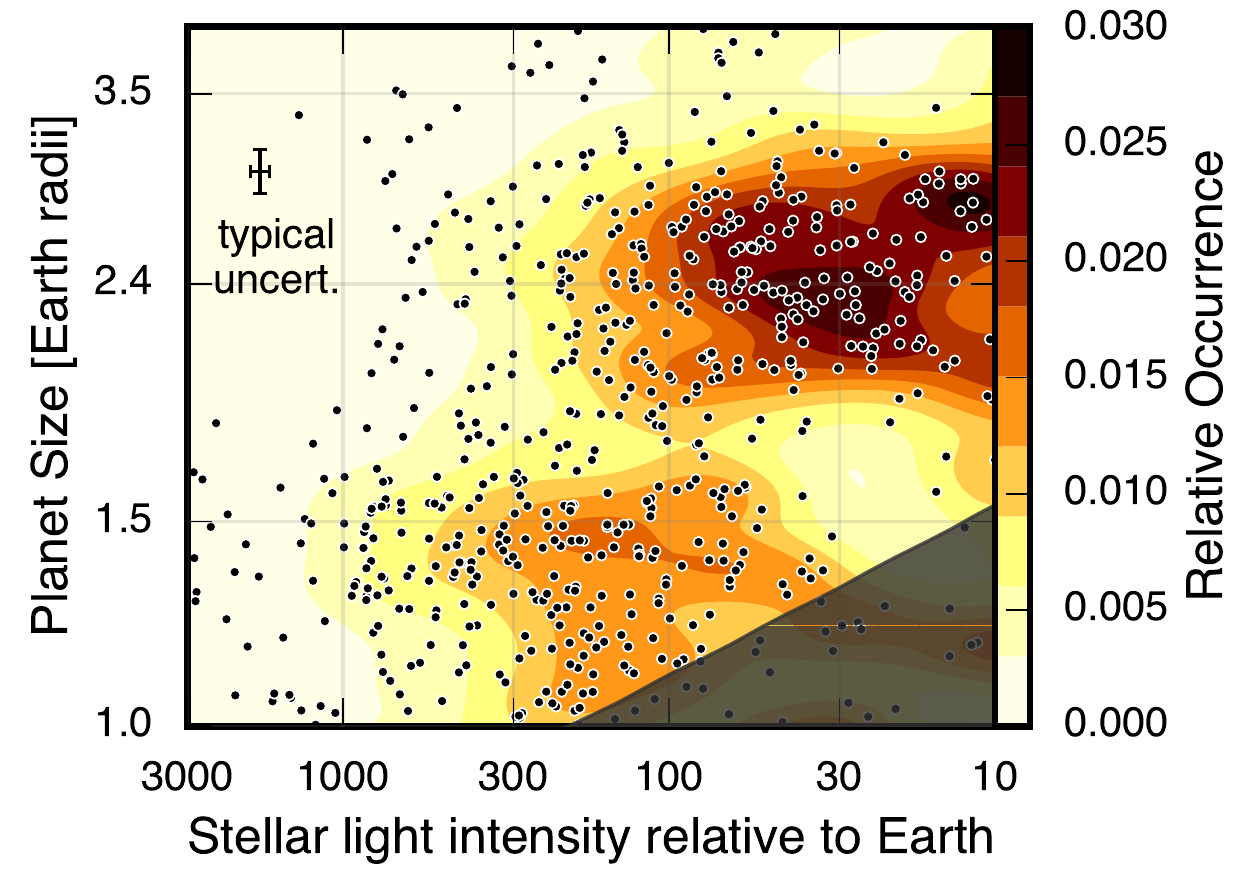}
\caption{{\em Left:} Two-dimensional distribution of planet size and orbital period. The median uncertainty is plotted in the upper left. {\em Right:} same as {\em left} but with insolation flux on the horizontal axis. In both plots, the two peaks in the population as observed by F17 are clearly visible, but with greater fidelity. \label{fig:insol-contour}}
\end{figure*}

In this work, we restricted our occurrence analysis to domains of $P$, \Rp, and \Sinc where pipeline completeness exceeds 25\% for our sample of \stat{cks-gaia-star-count} stars. Regions that do not meet this threshold are shaded in \ref{fig:radius-hist} and shown as gray triangles in Figure \ref{fig:insol-contour}. We placed an upper bound on the bias introduced by the IDEM estimator, by considering several regions abutting our completeness cutoff. We used the SysSim code from \citet{Hsu18} to estimate the occurrence using ABC. All values were consistent to within 1 $\sigma$. 

Figure \ref{fig:radius-hist} shows the radius distribution of close-in planets, i.e. the number of planets per star with orbital periods less than 100~days. Despite the increased precision relative to F17, the one-dimensional distribution of planet sizes is qualitatively the same. We confirm the presence of a gap in the occurrence distribution of planet radii at 1.5--2.0~\Re, as seen in F17 and several subsequent works (see, e.g., \citealt{VanEylen17}). The relative heights of the two peaks are similar indicating that the frequency of super-Earths and sub-Neptunes are nearly equal over the full period range analyzed in this work (0-100 days). We do not resolve additional small scale structure in the radius distribution, and the depth of the gap relative to the sub-Neptune and super-Earths peaks remains largely unchanged. This suggests that we are resolving the astrophysical scatter in the radii of the super-Earth and sub-Neptune populations, and that the gap is not completely devoid of planets. We quantify this astrophysical scatter in Section~\ref{sec:width}.

Figure \ref{fig:insol-contour} shows the two-dimensional occurrence distribution of planet radii as a function of orbital period and insolation flux. The contours show the relative planet occurrence computed using the weighted kernel density estimator (wKDE) described in F17. We used a Gaussian kernel spanning 40\% in \Sinc and 40\% $P$ and 5\% in \Rp. Smaller kernels offer higher resolution, but noisier contours due to few detected planets. Readers may create their own occurrence contours using the weights provided in Table~\ref{tab:weights}.

The radius gap is wider and more empty at $P \gtrsim 30$~days and $\Sinc \lesssim 50$~\Se. While there was tentative evidence for this in F17, the smaller \Rp uncertainties lend more confidence to this observation. The gap also appears to slope downward with increasing orbital period, which is consistent with the observations of \citet{VanEylen17} and with several theoretical models discussed in Section~\ref{sec:conclusions}.

\subsection{Intrinsic Spread in the Sizes of Super-Earths and Sub-Neptunes}
\label{sec:width}
Here, we consider the intrinsic astrophysical spread in the population of super-Earths and sub-Neptune planets and whether planets that appear to reside in the gap spanning 1.5 to 2.0~\Re could be explained by measurement uncertainties alone. Previous studies have struggled to measure the occurrence of planets over this narrow region of planet size. In F17, planet radius uncertainties were marginally smaller than the width of the gap, and the \cite{VanEylen17} asteroseismic analysis suffered from a small number of detected planets.

\begin{figure*}[ht]
\centering
\includegraphics[width=1\textwidth]{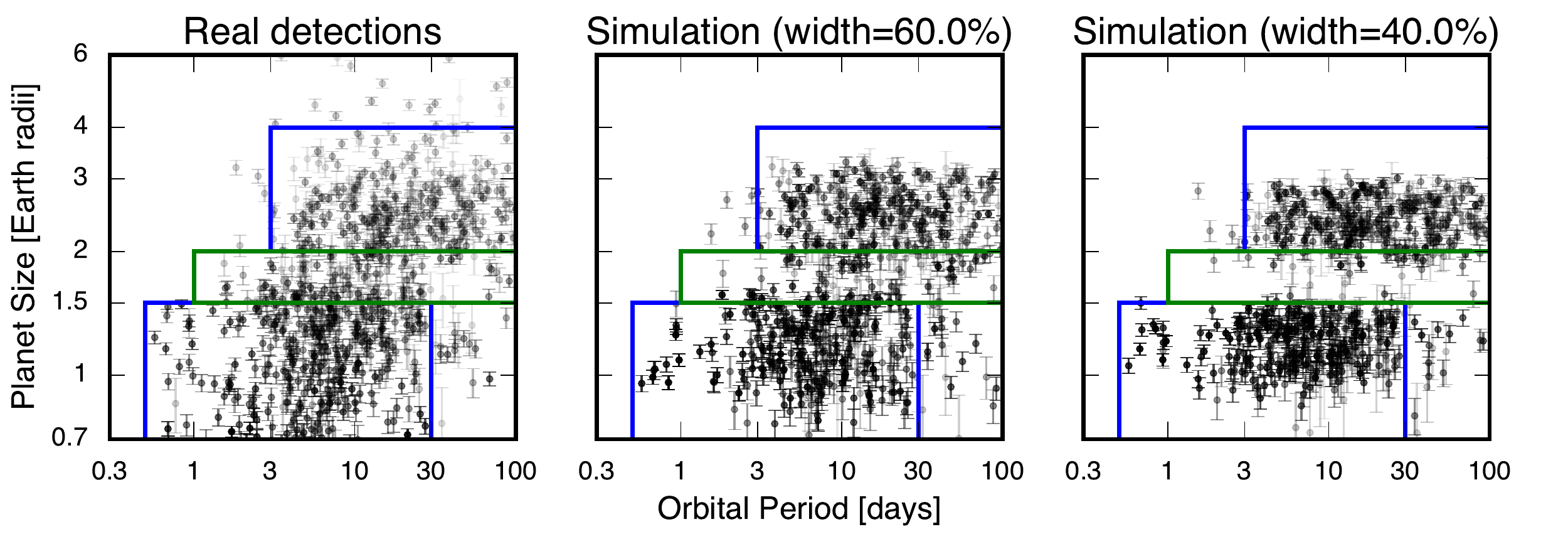}
\caption{Toy model demonstrating that the two populations of planets have intrinsic widths. \emph{Left:} Real planet detections with boxes demarking the boundaries defined for the population of large planets (\Rp = 2.0--4.0~\Re), small planets (\Rp = 0.7--1.5~\Re), and the gap between them (\Rp = 1.5--2.0~\Re). We find that the data is well-described by two populations with a 60\% intrinsic spread in their radii (\emph{middle}). Decreasing that width to 40\% is a clear mismatch to the data (\emph{right}). Our toy model is described in Section~\ref{sec:width}.
\label{fig:width}}
\end{figure*}

We show the filtered distribution of planets in Figure~\ref{fig:width}, and we identify two fiducial planet classes, ``super-Earths'' and ``sub-Neptunes,'' as well as the radius gap. The large number of planets (\stat{planet-filtered-gap-count}) residing within the gap appear to be inconsistent with scatter from above and below. To test this, we constructed a toy model to assess whether they could be explained by measurement uncertainties alone.

In or toy model, we took the observed planet detections and assigned them to one of the two planet classes, based on whether they resided in one of the two boxes shown in Figure~\ref{fig:width}. For each super-Earth and sub-Neptune we assigned a new radius from uniform distributions with centers at 1.2~\Re or 2.4~\Re, respectively, which correspond to the locations of the observed peaks in the radius distribution. The orbital periods were retained from the actual detections. The fractional width shared by both distributions $W$ is a free parameter in this model. 

We simulated planet detections over a range of $W$ and computed a figure of merit $\mathrm{FOM} = \sum_i (N_{\mathrm{real},i} - N_{\mathrm{sim},i})^{2}$, where $N$ is the number of detections within box $i$ (plotted in Figure~\ref{fig:width}).

The FOM and visual inspection identified an intrinsic spread of $W = 60\%$ as a good match to the data (see Figure~\ref{fig:width}). In our best-fitting toy model, the super-Earth and sub-Neptune populations span 0.85 to 1.55~\Re and 1.7 to 3.1~\Re, respectively; it is inconsistent with a population devoid of planets between 1.5 and 2.0~\Re. 

As a limiting case, we show a model with $W = 40\%$ in Figure~\ref{fig:width}. Here, the super-Earth and sub-Neptune population span 0.96 to 1.44~\Re and 1.92 to 2.88~\Re, respectively, which approximate the upper and lower boundaries of the gap. This model produces a much emptier gap and is an obvious mismatch with the observations.

We recognize that this toy model does not capture the detailed radius distribution of planets, most notably the tail of planets larger than 3~\Re. A more detailed study might use a different distribution to model the planet radii, such as a Gaussian, Rayleigh, or a non-parametric distribution. Nonetheless, we have constrained the intrinsic dispersion in the size of the super-Earth and sub-Neptune populations to be $\approx$60\%. While there is a dip in the occurrence of close-in planets from 1.5 to 2.0~\Re, occurrence does not fall to zero. These interpretations were not possible in previous studies, due to larger radius uncertainties or limited numbers of detected planets. 

\subsection{Trends with Host-Star Mass}
\begin{figure*}[ht]
\centering
\includegraphics[width=0.8\textwidth]{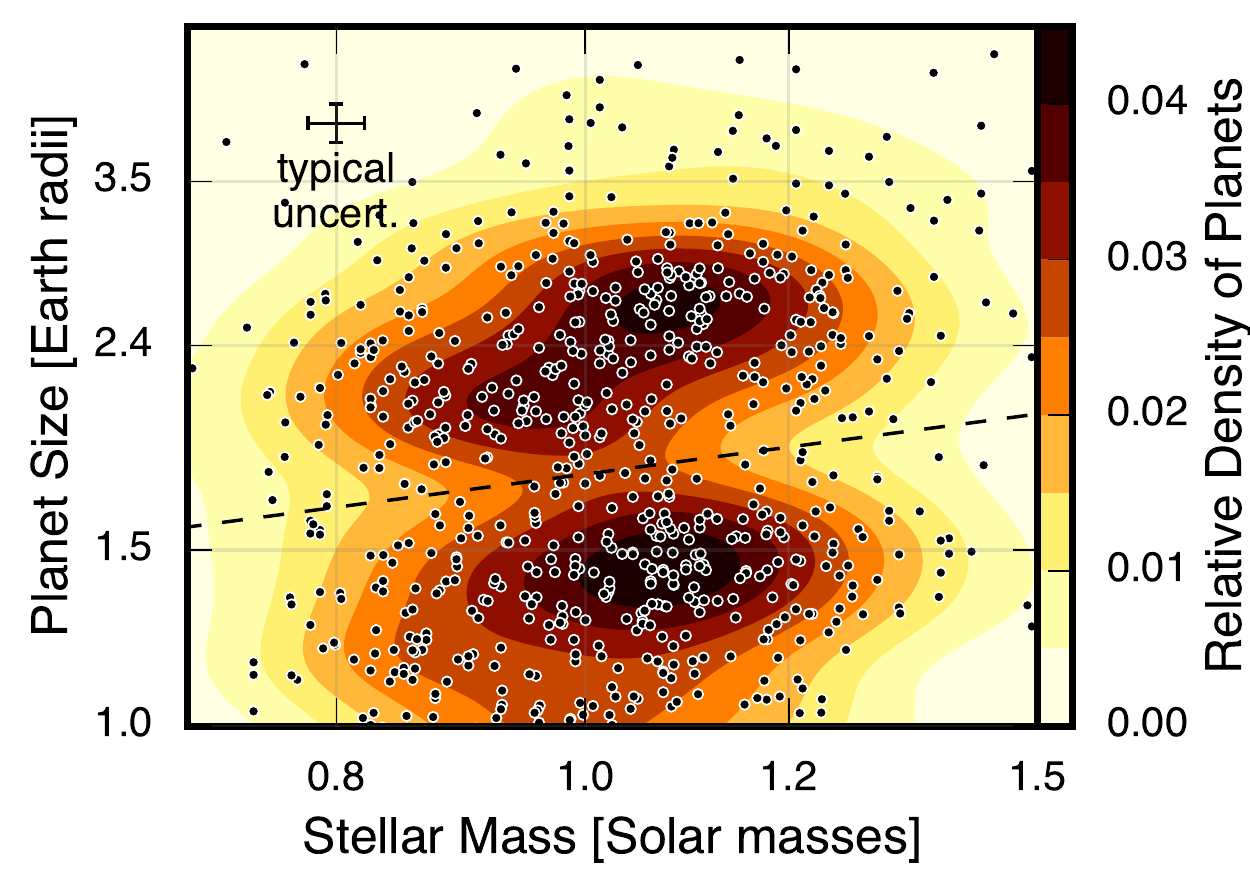}
\caption{Two-dimensional distribution of stellar mass and planet size. The median uncertainty is plotted in the upper left. As we see in Figure \ref{fig:insol-contour-masscuts}, the position of the gap, and the population of planets on either side of the gap increases monotonically with increasing stellar mass. We plot a dashed line at the location of the gap to guide the eye. For stars with masses of $\approx$0.8 \Msun the gap falls at $\approx$1.6 \Re, while for host stars with masses of $\approx$1.2 \Msun the gap occurs at $\approx$2.0 \Re. The peaks of the two populations of planets on either side of the gap also shift in the same way. Planets smaller than 4 \Re tend to be larger around more massive stars and the same is true for the gap between the two populations. The relative occurrence rate between the two populations remains constant for all stellar masses analyzed in this work.}
\label{fig:smass-contour}
\end{figure*}

\begin{figure*}[ht]
\centering
\includegraphics[width=1\textwidth]{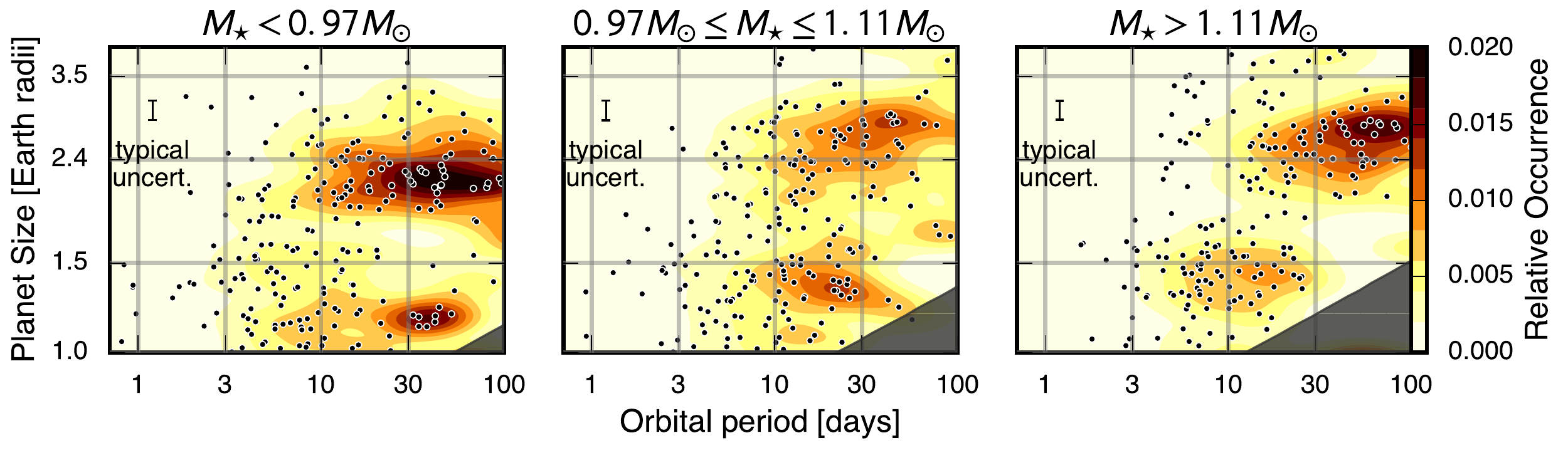}
\includegraphics[width=1\textwidth]{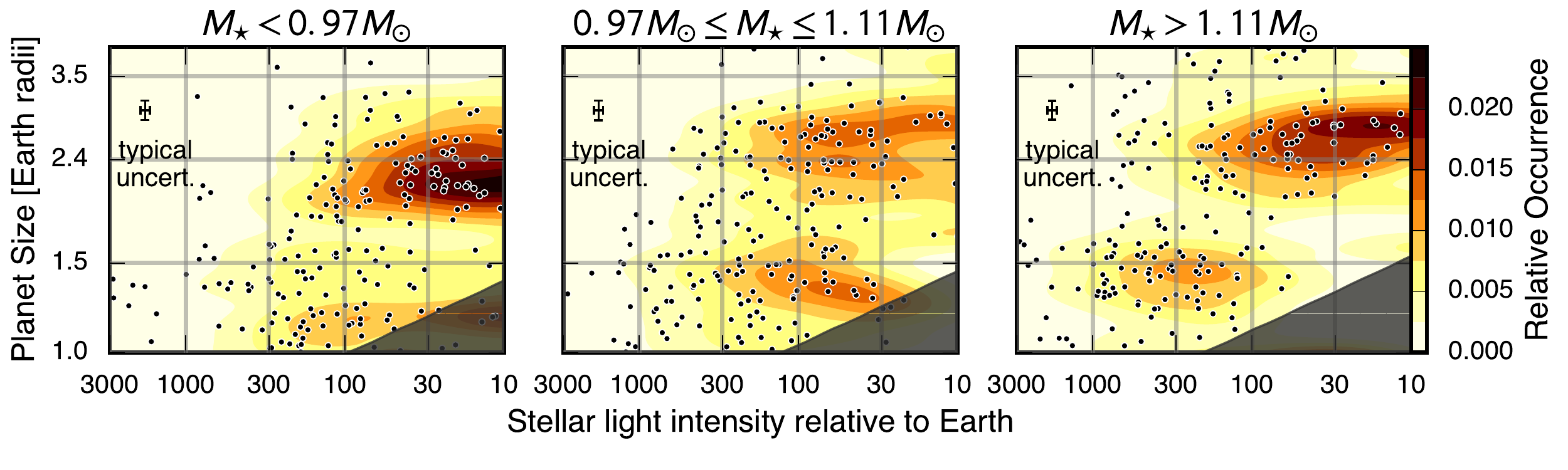}
\caption{{\em Top row:} the two-dimensional distribution of planet size and orbital period for three bins of stellar mass. The typical size of super-Earths (\Rp = 1.0--1.7~\Re) and sub-Neptunes (\Rp = 1.7--4.0~\Re) increases with stellar mass while typical orbital periods are roughly constant. {\em Bottom row:} same as {\em top row}, but with insolation flux on the horizontal axis. The population of super-Earths and sub-Neptunes shifts to higher incident flux for higher mass stars.}
\label{fig:insol-contour-masscuts}
\end{figure*}

We plot planet size vs. stellar mass in Figure \ref{fig:smass-contour} in order to investigate potential changes in the structure of the planet radius distribution as a function of stellar host mass. This Figure shows that the transition radius between the two populations increases monotonically with stellar mass. The gap occurs near 1.6 \Re for planets orbiting host stars with masses near 0.8 \Msun and moves to $\approx$2.0 \Re for planets orbiting stars with masses above 1.2 \Msun. We also split the sample into three bins of stellar mass: \Mstar $<$ 0.96~\Msun, \Mstar = 0.96--1.11~\Msun, and \Mstar $>$ 1.11~\Msun. We chose bin boundaries such that the three bins captured equal numbers of planets. Figure~\ref{fig:insol-contour-masscuts} shows the planet population in the $P$-\Rp and \Sinc-\Rp planes for each of the three mass bins. The gap is clearly visible in each of the three stellar mass bins, and appears wider than the gap from the combined sample shown in Figure~\ref{fig:insol-contour}. 

We observe several trends with stellar mass. First, the typical size of super-Earth and sub-Neptune planets increases with increasing stellar mass, an observation that we quantify later in this section. This explains why the planet populations are better separated when one considers a narrow range of stellar mass; when all three mass groups are combined the distributions overlap. It also helps to clarify why the planet populations in \cite{VanEylen17} seemed to be more separated compared to those in F17. The asteroseismic sample was heavily weighted toward stars more massive than the sun, and is more directly comparable to the $P$--\Rp distribution of our high mass bin. The top right panel of our Figure~\ref{fig:insol-contour-masscuts} is a closer match to Figure~2 from \cite{VanEylen17} than the upper left panel of our Figure~\ref{fig:insol-contour-masscuts}. 

\begin{figure*}
\centering
\includegraphics[width=0.8\textwidth]{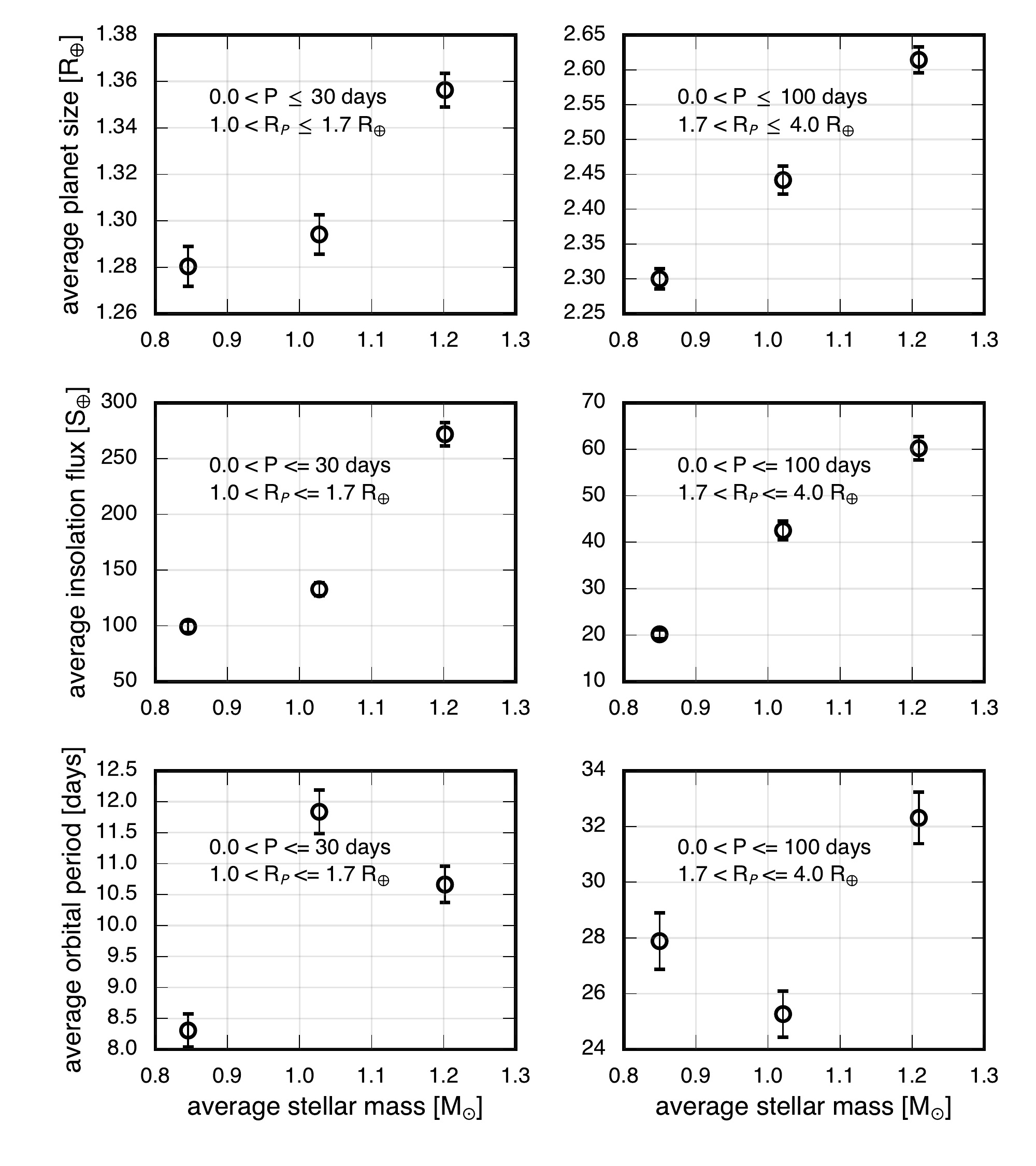}
\caption{Mean planet properties as a function of mean stellar mass for super-Earths and  sub-Neptunes (left and right columns respectively). The top, middle, and bottom rows show the weighted average of planet radius, insolation flux, and orbital period, respectively. Planets around more massive stars tend to be larger and hotter than those around lower mass stars, but their orbital periods are similar.}
\label{fig:mean-values}
\end{figure*}

To quantify the change in typical planet size with stellar mass, we calculated the mean planet radius for sub-Neptunes (1.7--4.0~\Re, and $P<100$ days) and super-Earths (1.0--1.7 \Re, and $P<30$ days). We weighted each radius by the $w_i$ weights used in the occurrence calculations, described in Section~\ref{sec:occurrence}. Figure \ref{fig:mean-values} shows these mean planet parameters as a function of stellar mass. Consistent with visual inspection of Figure~\ref{fig:insol-contour-masscuts}, we see monotonically increasing planet size with increasing stellar host star mass in both the super-Earth and sub-Neptune planets. 

Although the trend with stellar mass is strong, we caution that stellar metallicity may be a confounding factor. More massive stars are younger on average and are thus more metal-rich due to galactic chemical enrichment. Indeed, \cite{Petigura18b} observed a correlation between planet size and host star metallicity in the CKS sample and this correlation has been observed previously in many different samples \citep[e.g.][]{Santos04a, Fischer05, Sousa08, Ghezzi10, Buchhave14, Schlaufman15, Wang15}. The solid component of the protoplanetary disk likely tracks both stellar metallicity and stellar mass. Therefore, we expect planet size to be correlated with both stellar mass and metallicity. Future studies spanning a larger range of stellar mass and metallicity are necessary to resolve this ambiguity.

Previous studies have noted a desert of highly-irradiated sub-Neptune planets (see, e.g., \citealt{Lundkvist16} and \citealt{Mazeh16}). We observe this sub-Neptune desert in our three mass bins (Figure~\ref{fig:insol-contour-masscuts}), but note that it shifts to higher incident stellar flux around high mass stars. This trend is highly significant. Figure~\ref{fig:mean-values} shows the average \Sinc as function \Mstar. The mean \Sinc for both the super-Earths and sub-Neptunes increases by $3\times$ over a relatively narrow range of average \Mstar, 0.85~\Msun to 1.2~\Msun. One explanation could be that the orbital periods of small planets decreases with stellar mass. However, the mean orbital periods for the three different mass bins are consistent to $\approx$30\%. 

\begin{figure}[ht]
\centering
\includegraphics[width=0.49\textwidth]{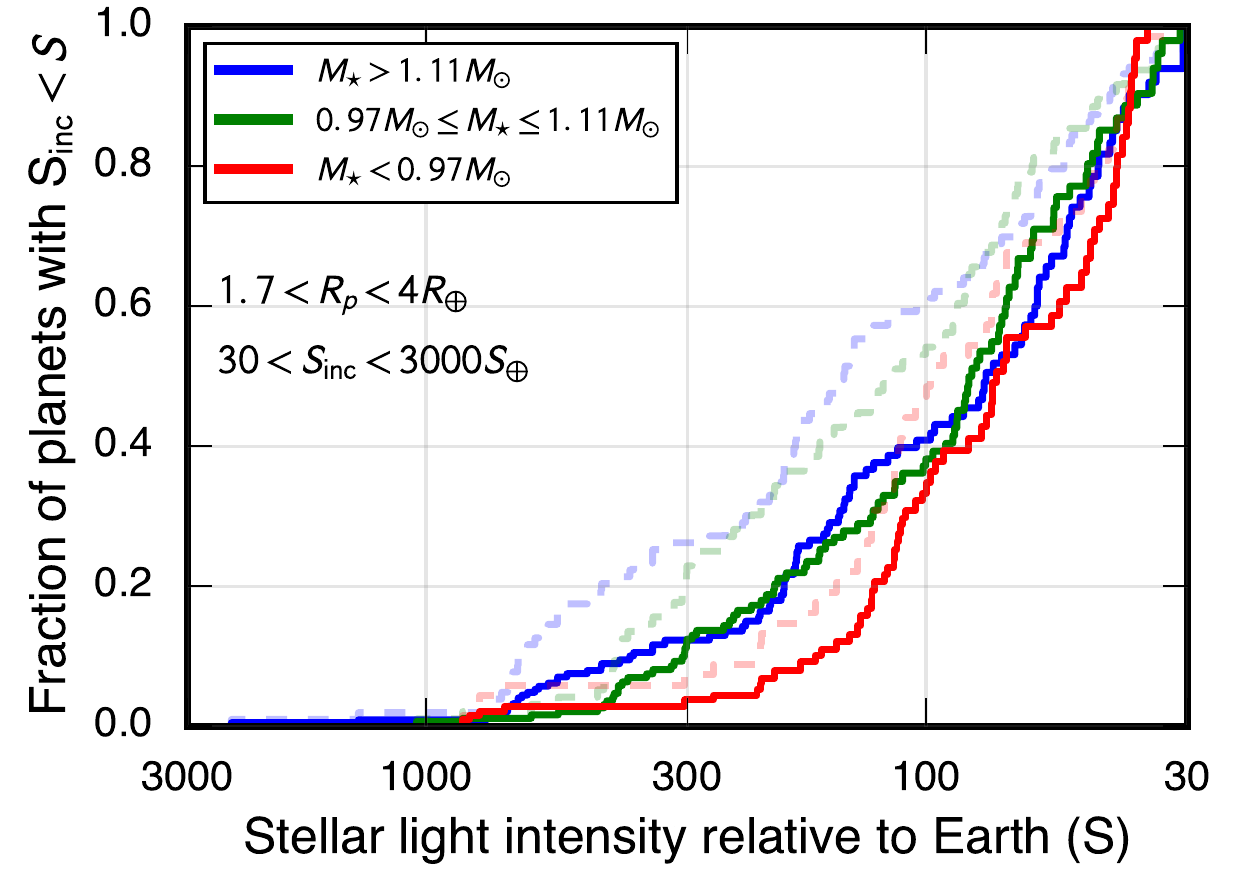}
\caption{Cumulative distribution of planets as a function of insolation flux. The solid lines are corrected for completeness and the dashed lines are not corrected. Planets residing in very high insolation flux environments tend to be more rare around low mass stars. For example, $\sim$20\% of the planet population orbiting stars more massive than 1.11 \msun have \Sinc $>$ 200 \Se, compared to only $\sim$10\% of planets orbiting stars less massive than 0.96~\Msun.}
\label{fig:desert-edge}
\end{figure}

Figure~\ref{fig:desert-edge} shows the cumulative fraction of hot (\Sinc = 30--3000~\Se) sub-Neptune size planets (\Rp = 1.7--4.0~\Re) as a function of insolation flux, highlighting the shift of the $\Sinc$ sub-Neptune desert with stellar mass. In the high stellar mass sample, 10\% of planets have \Sinc $>$ 300~\Se. For the low stellar mass sample, one must include planets out to 150~\Sinc to encompass 10\% of the population.

\section{Discussion and Conclusions}
\label{sec:conclusions}
We analyzed the \Kepler planet population after improving the radius precision of the stellar hosts and their planets. This improvement leveraged both CKS spectroscopy and \Gaia DR2 parallaxes. Our median stellar radius precision is now 2\% compared to 11\% in J17. The uncertainty in our planet radii are now typically 5\% and are limited by uncertainties in the \Kepler transit modeling, rather than the stellar radius uncertainties as in J17.

With these improved planet radii, we examined the population of small planets at higher resolution. We confirmed the existence of the F17 radius gap between 1.5 and 2.0~\Re, with more precise and independently derived planet radii. The overall radius distribution is similar to that of F17, which demonstrates that we are resolving the intrinsic spread of the super-Earth and sub-Neptune populations, which span $\approx$60\% in radius. We also demonstrated that the gap from 1.5 to 2.0~\Re is not devoid of planets, a conclusion previously obscured by measurement uncertainties or small sample sizes. We observed a correlation of the average planet size and average insolation flux with stellar host mass. However, there is no significant correlation in the average orbital period as a function of average stellar host mass.

Here, we interpret our findings in the context of two theories that have been proposed to explain the distribution planets between the size of Earth and Neptune:
\begin{enumerate}
\item {\em Mass loss by photoevaporation}. In this mechanism, X-ray and UV radiation heats the outer layers of a planet's envelope and drives mass loss. Several groups considered photoevaporation and predicted the planet radius gap before it was observed in F17, including \cite{Owen13}, \cite{Lopez13b}, \cite{Jin14}, and \cite{Chen16}. Following F17, \cite{Owen17} developed additional analytic photoevaporation theory and performed a population synthesis analysis comparing their simulated populations to the F17 occurrence measurements.

\item {\em Core-powered mass loss}. In this mechanism, luminosity from a cooling rocky core heats a planet's envelope and drives mass loss. \cite{Ginzburg16} developed the theory of core-powered mass loss and computed mass loss rates. \cite{Ginzburg18} performed a population synthesis with comparisons to the F17 radius distribution and demonstrated that core-powered mass loss could explain the bimodal radius distribution.
\end{enumerate}


Both theories can explain a bimodal population of planet sizes composed of two subpopulations: a population of bare rocky cores and a population with H/He envelopes with mass fractions of a few percent. Because both mass loss mechanisms are more efficient at high levels of incident stellar flux, they both predict that the population of sub-Neptunes should be offset to lower insolation fluxes compared to the super-Earths 

A key difference between the two mechanisms is the expected dependence on stellar mass. Core-powered mass loss depends only on properties of the planet and {\em bolometric incident stellar flux}. All else being equal, this mechanism predicts no dependence of the planet population as a function of \Mstar. In contrast, the efficiency of photoevaporation depends on the {\em time-integrated XUV flux}, or ``fluence.'' This quantity is a strong function of stellar mass since $\int (L_{X} / L_{\mathrm{bol}} ) d t \propto \Mstar^{-3}$ \citep{Jackson12}. Therefore, photoevaporation predicts that the population of sub-Neptunes should shift to lower $\Sinc$ with decreasing stellar mass, due to increased activity around lower mass stars. The shifts in the \Sinc-\Rp distribution of planets with \Mstar are consistent with this prediction from photoevaporation.

The lack of a strong $P$--\Mstar dependence is also consistent with photoevaporation. \cite{Owen17} showed that the mass loss timescale $t = M / \dot{M} \propto P^{1.4} \Mstar^{-0.48} \propto \Sinc^{1.06} \Mstar^{2.2}$. Photoevaporation thus has a steeper dependence on \Mstar at fixed \Sinc than at fixed $P$. This naturally explains why we see a strong trend in planet $\Sinc$ with stellar mass and no significant trend with $P$ in Figure~\ref{fig:mean-values}.

Other super-Earth formation mechanisms have been proposed that could potentially produce a gap in the size distribution including delayed formation in a gas-poor disk \cite[e.g.][]{Lee14, Lee16}, and sculpting by giant impacts \cite[e.g.][]{Liu15, Schlichting15, Inamdar16}. Formation in a gas-poor disk without any sculpting by photoevaporation would produce a gap radius that does not change with orbital period and is inconsistent with the results of \citet{VanEylen17}. Sculpting by giant impacts alone predicts that the gap radius would be found at larger radii at longer orbital periods \citep{Lopez16} which is the opposite of the trend found by \citet{VanEylen17}.

We interpret the observed stellar mass dependence of the planet population as evidence supporting  \mbox{photoevaporation} model. However, these mechanisms are not mutually exclusive and all of them could be operating simultaneously or or at different times during the formation of planetary systems. If photoevaporation is the dominate mechanism for sculpting planet envelopes, one may fit the observed distribution of planets to constrain important quantities like the distribution of planet core masses, envelope fractions, and core compositions \citep{Owen17}.

Due to the magnitude-limited nature of CKS, our analysis was restricted to a fairly narrow range in \Mstar, spanning 0.85 to 1.2~\Msun.  Previous studies of the radius distribution of planets orbiting M dwarfs have shown that these planets tend to be smaller on average (\citealt{Morton14,Dressing15}). This may be an extension of the stellar mass dependence of the planet population observed in this work. However, no study of planets orbiting low-mass stars to date has detected a gap in the radius distribution. Such a detection (or lack thereof) would reveal insights into the structure and formation of planets around low-mass stars. This motivates future high precision studies of large samples of planets orbiting K and M dwarfs. Such studies would provide additional leverage on \Mstar to test the dependence of the planet population on stellar mass and to constrain the mechanisms that form and sculpt planets.

\facility{Keck:I (HIRES), \Kepler, \Gaia}

\acknowledgments{We thank the \Kepler and \Gaia teams for years of work making these precious datasets possible. We are grateful for Andrew Howard's guidance and comments on our manuscript. We thank Eddie Schlafly and Gregory Green for guidance regarding the treatment of dust extinction. We thank Alberto Krone-Martins for discussions regarding the \Gaia mission and its data products. Daniel Huber kindly assisted with the isoclassify package. We thank D. Hsu and E. Ford for help using their SysSym code.

EAP acknowledges support from Hubble Fellowship grant HST-HF2-51365.001-A awarded by the Space Telescope Science Institute, which is operated by the Association of Universities for Research in Astronomy, Inc. for NASA under contract NAS 5-26555. 

This research has made use of NASA's Astrophysics Data System

This work has made use of data from the European Space Agency (ESA) mission
{\it Gaia} (\url{https://www.cosmos.esa.int/gaia}), processed by the {\it Gaia}
Data Processing and Analysis Consortium (DPAC,
\url{https://www.cosmos.esa.int/web/gaia/dpac/consortium}). Funding for the DPAC
has been provided by national institutions, in particular the institutions
participating in the {\it Gaia} Multilateral Agreement.

Finally, the authors wish to recognize and acknowledge the very significant cultural role and reverence that the summit of Maunakea has long had within the indigenous Hawaiian community.  We are most fortunate to have the opportunity to conduct observations from this mountain.}

\software{All code used in this paper is available at \url{https://github.com/California-Planet-Search/cksgaia/}. We made use of the following publicly available Python modules: astropy \citep{Astropy-Collaboration13}, isoclassify \citep{Huber17}, lmfit \citep{Newville14}, matplotlib \citep{Hunter07}, numpy/scipy \citep{numpy/scipy}, and pandas \citep{pandas}.}

\bibliographystyle{aasjournal}
\bibliography{references}

\clearpage

\begin{deluxetable*}{lRRRRRRRRRRRRR}
\tablecaption{Stellar Properties\label{tab:star}}
\tabletypesize{\scriptsize}
\tablecolumns{12}
\tablewidth{0pt}
\tablehead{
	\colhead{KOI} & 
	\colhead{\teff} &
	\colhead{\fe} &
	\colhead{$m_K$} &
    \colhead{$\pi$} &
    \colhead{$R$} & 
	\colhead{$M_{\mathrm{iso}}$} & 
	\colhead{$R_{\mathrm{iso}}$} & 
	\colhead{$\rho_{\mathrm{iso}}$} & 
    \colhead{age$_{\mathrm{iso}}$} &
    \colhead{$\pi_{\mathrm{spec}}$} &
    \colhead{$r_8$} &
	\colhead{RCF} 
    \\
    \colhead{} & 
	\colhead{K} &
	\colhead{dex} & 
    \colhead{mag} & 
    \colhead{mas} &
    \colhead{\Rsun} & 
	\colhead{\Msun} & 
	\colhead{\Rsun} & 
	\colhead{g/cc} & 
    \colhead{dex} & 
    \colhead{mas} &
    \colhead{} &
    \colhead{} 
}
\startdata
\input{tab_star-stub.tex}
\enddata
\tablecomments{Properties of planet hosting stars. \teff and \fe are from P17, $m_K$ is the $K$-band apparent magnitude from 2MASS (\citealt{Skrutskie06}, see Section~\ref{sec:photometry}), $\pi$ is the trigonometric parallax from \Gaia DR2 (\citealt{Gaia18}, see Section~\ref{sec:parallax}). The following quantities are described in Section~\ref{sec:detailed-modeling}: $R$ is the adopted stellar radius, computed using the Stefan-Boltzmann law; stellar properties with the `iso' subscript incorporate constraints from the MIST isochrones; and $\pi_\mathrm{spec}$ is the ``spectroscopic parallax.'' $r_8$ encodes contaminating flux from neighboring stars within 8~arcsec in $G$-band (see Section~\ref{sec:occurrence}). RCF is the ``radius correction factor'' computed by \cite{Furlan17} (see Section~\ref{sec:occurrence}). Table \ref{tab:star} is published in its entirety in machine-readable format. A portion is shown here for guidance regarding its form and content.}
\end{deluxetable*}

\begin{deluxetable*}{lRRRRRr}
\tablecaption{Planet Properties\label{tab:planet}}
\tabletypesize{\scriptsize}
\tablecolumns{6}
\tablewidth{0pt}
\tablehead{
	\colhead{Planet} & 
	\colhead{$P$} &
	\colhead{\Rp/\Rstar} & 
	\colhead{\Rp} & 
	\colhead{$a$} &
	\colhead{\Sinc} &
    \\
    \colhead{candidate} & 
	\colhead{d} &
	\colhead{} &
	\colhead{\Re} & 
	\colhead{AU} & 
	\colhead{\Se} &
}
\startdata
\input{tab_planet-stub.tex}
\enddata
\tablecomments{Planetary properties. Period $P$ and planet-to-star radius ratio $\Rp/\Rstar$ are from \cite{Mullally15}. Planet size \Rp, semi-major axis $a$, and incident stellar flux relative to Earth \Sinc are derived from the updated stellar properties in Table~\ref{tab:star}.
Table \ref{tab:star} is published in its entirety in machine-readable format with full numerical precision and uncertainties. A portion is shown here for guidance regarding its form and content.}
\end{deluxetable*}

\begin{deluxetable*}{lrccr}
\tablecaption{Planet Detection Statistics\label{tab:weights}}
\tabletypesize{\scriptsize}
\tablewidth{0pt}
\tablehead{
	\colhead{Planet} & 
	\colhead{SNR} &
	\colhead{Detection probability} & 
    \colhead{Transit probability} &
    \colhead{Weight}
    \\
    \colhead{candidate} & 
	\colhead{$m_{i}$} &
	\colhead{$p_{\rm det}$} & 
    \colhead{$p_{\rm tr}$} &
    \colhead{$1/w_i$}
}
\startdata
\input{tab_weight-tex-stub.tex}
\enddata
\tablecomments{Table \ref{tab:weights} is available in its entirety in machine-readable format. A portion is shown here for guidance regarding its form and content.
This table contains only the subset of planet detections that passed the filters described in Section \ref{sec:planet-population}.}
\end{deluxetable*}

\end{document}

%% file: newcommand_gen.tex
\newcommand{\msun}{M$_{\odot}~$}

\newcommand{\Kepler}{\textit{Kepler}\xspace} 
\newcommand{\Kp}{\textit{Kp}\xspace}

\newcommand{\Gaia}{\textit{Gaia}\xspace}

\newcommand{\Mstar}{\ensuremath{M_{\star}}\xspace}
\newcommand{\Rstar}{\ensuremath{R_{\star}}\xspace} 
 
\newcommand{\teff}{\ensuremath{T_{\mathrm{eff}}}\xspace}  
\newcommand{\logg}{\ensuremath{\log g}\xspace} 
\newcommand{\fe}{[Fe/H]\xspace}

\newcommand{\Lbol}{\ensuremath{L_{\mathrm{bol}}}\xspace}

\newcommand{\Mbol}{\ensuremath{M_{\mathrm{bol}}}\xspace}

\newcommand{\Rp}{\ensuremath{R_p}\xspace}

\newcommand{\Sinc}{\ensuremath{S_{\mathrm{inc}}}\xspace}

\newcommand{\Se}{\ensuremath{S_{\oplus}}\xspace}
 
\renewcommand{\Re}{\ensuremath{R_{\oplus}}\xspace}

\newcommand{\Rsun}{\ensuremath{R_{\odot}}\xspace }
\newcommand{\Msun}{\ensuremath{M_{\odot}}\xspace}
 
\newcommand{\K}{\ensuremath{\mathrm{K}}\xspace}

\newcommand{\sigmasb}{\ensuremath{\sigma_\mathrm{sb}}\xspace}

%% file: values.tex
\usepackage{xstring} 

\newcommand{\val}[1]{%
  \IfEqCase{#1}{%
{nstars-cks}{ 1305 }
{ncand-cks}{2025}
{med-srad-frac-err}{11\%}
{med-smass-frac-err}{ 4\% }
{cks-rp-frac-err-median}{12\%}
{gamma-k}{$k=17.56$}
{gamma-l}{$l=1.00$ (fixed)}
{gamma-theta}{$\theta=0.49$}
{logistic-b}{$b=9.54$}
{logistic-c}{$c=10.09$}
{logistic-d}{$d=0.90$}
{num-sim-planets}{45000}
{num-planets-filtered}{900}
{num-stars-inj}{3840}
{error-bin-fudge}{1.5--2.9}
{nstars-occ}{36,075}
{ncand-sample}{900}
{occ-ratio}{$0.8 \pm 0.2$}
{kstest}{0.003}
{adtest}{0.012}
{tr-prob-sm}{6\%}
{tr-prob-lg}{3\%}
{det-prob-sm}{86\%}
{det-prob-lg}{96\%}
}[\PackageError{tree}{Undefined option to tree: #1}{}]%
}%

%% file: val_stat.tex
\newcommand{\stat}[1]{
\IfEqCase{#1}{%
{cks-star-count}{1279}%
{cks-mag-star-count}{1279}%
{cks-gaia-star-xmatch-count}{1257}%
{cks-gaia-star-count}{1189}%
{cks-gaia-sparallax-med}{1.5}%
{cks-gaia-sparallax-ferr-med}{1.3}%
{cks-gaia-distmod-err-med}{0.01}%
{cks-gaia-distmod-med}{9.26}%
{ak-med}{0.011}%
{ak-min}{0.001}%
{ak-max}{0.054}%
{ak-med-err}{0.004}%
{ebv-med}{0.048}%
{ebv-med-err}{0.007}%
{mk-med}{12.24}%
{mk-err-med}{0.02}%
{steff-med}{5698}%
{ror-med}{1.7}%
{ror-ferr-med}{4.1}%
{srad-med}{1.1}%
{srad-ferr-med}{2.7}%
{prad-ferr-med}{5.2}%
{prad-med}{2.1}%
{cks-s15-count}{29}%
{cks-s15-srad-mean}{0.9\% larger}%
{cks-s15-srad-std}{1.9\%}%
{cks-h13-count}{63}%
{cks-h13-srad-mean}{0.03\% smaller}%
{cks-h13-srad-std}{3.5\%}%
{cks-j17-count}{1050}%
{cks-j17-srad-mean}{0.7\% larger}%
{cks-j17-srad-std}{13.9\%}%
{cks-gaia2-count}{1077}%
{cks-gaia2-srad-err-mean}{6.9\%}%
{cks-gaia2-srad-mean}{3.0\% larger}%
{cks-gaia2-srad-std}{6.2\%}%
{num-planets-filtered}{907}%
{planet-count}{1901}%
{star-filtered-count}{24981}%
{planet-filtered-count}{907}%
{planet-filtered-gap-count}{129}%
}[XX]%
}

%% file: tab_star-stub.tex
K00001 & 5819 & 0.01 & 9.8 & 4.67 & 1.04 & 0.99 & 1.04 & 0.87 & 9.74 & 4.76 & 1.00 & 1.010 \\
K00002 & 6449 & 0.20 & 9.3 & 2.96 & 1.97 & 1.51 & 1.96 & 0.20 & 9.25 & 3.66 & 1.00 & 1.003 \\
K00006 & 6348 & 0.04 & 11.0 & 2.13 & 1.30 & 1.20 & 1.28 & 0.57 & 9.32 & 2.20 & 1.01 & 1.001 \\
K00007 & 5827 & 0.18 & 10.8 & 2.07 & 1.51 & 1.15 & 1.51 & 0.34 & 9.78 & 2.11 & 1.00 & \nodata \\
K00008 & 5891 & -0.07 & 11.0 & 3.01 & 0.94 & 1.00 & 0.93 & 1.23 & 9.18 & 2.90 & 1.00 & \nodata \\
K00010 & 6181 & -0.08 & 12.3 & 1.00 & 1.54 & 1.15 & 1.53 & 0.32 & 9.67 & 1.21 & 1.01 & 1.001 \\
K00017 & 5660 & 0.36 & 11.6 & 1.73 & 1.26 & 1.09 & 1.25 & 0.56 & 9.81 & 1.54 & 1.03 & \nodata \\
K00018 & 6332 & 0.02 & 11.8 & 1.14 & 1.74 & 1.31 & 1.69 & 0.27 & 9.46 & 1.25 & 1.02 & 1.005 \\
K00020 & 5926 & 0.03 & 12.1 & 1.16 & 1.50 & 1.09 & 1.50 & 0.32 & 9.83 & 1.10 & 1.00 & \nodata \\
K00022 & 5891 & 0.21 & 12.0 & 1.41 & 1.25 & 1.12 & 1.25 & 0.58 & 9.67 & 1.31 & 1.00 & \nodata \\
K00041 & 5854 & 0.10 & 9.8 & 3.31 & 1.53 & 1.10 & 1.52 & 0.31 & 9.84 & 3.41 & 1.00 & 1.008 \\
K00046 & 5661 & 0.39 & 12.0 & 1.10 & 1.66 & 1.24 & 1.64 & 0.28 & 9.72 & 1.19 & 1.00 & \nodata \\

%% file: tab_planet-stub.tex
K00001.01 & 2.5 & 0.124  & 14.14 & 0.036 & 882 \\  
K00002.01 & 2.2 & 0.075  & 16.25 & 0.038 & 4161 \\  
K00006.01 & 1.3 & 0.294  & 41.94 & 0.025 & 3852 \\  
K00007.01 & 3.2 & 0.025  & 4.08 & 0.045 & 1180 \\  
K00008.01 & 1.2 & 0.019  & 1.90 & 0.022 & 2031 \\  
K00010.01 & 3.5 & 0.094  & 15.70 & 0.047 & 1375 \\  
K00017.01 & 3.2 & 0.095  & 13.10 & 0.044 & 753 \\  
K00018.01 & 3.5 & 0.080  & 15.23 & 0.050 & 1759 \\  
K00020.01 & 4.4 & 0.118  & 19.38 & 0.054 & 848 \\  
K00022.01 & 7.9 & 0.094  & 12.85 & 0.081 & 260 \\  
K00041.01 & 12.8 & 0.014  & 2.34 & 0.111 & 201 \\  
K00041.02 & 6.9 & 0.008  & 1.34 & 0.073 & 459 \\  
K00041.03 & 35.3 & 0.009  & 1.54 & 0.217 & 52 \\  
K00046.01 & 3.5 & 0.033  & 5.97 & 0.048 & 1076 \\  
K00046.02 & 6.0 & 0.007  & 1.24 & 0.070 & 520 \\  

%% file: tab_weight-tex-stub.tex
 K00958.01 & 186.24 & 0.97 & 0.02 & 49.24 \\
 K04053.01 &  21.03 & 0.77 & 0.17 &  7.71 \\
 K04212.02 &   8.77 & 0.81 & 0.05 & 22.85 \\
 K04212.01 &  16.53 & 0.93 & 0.08 & 13.79 \\
 K01001.01 &  37.27 & 0.99 & 0.03 & 32.14 \\
 K01001.02 &  15.49 & 0.96 & 0.01 & 75.81 \\
 K02534.01 &  22.64 & 0.94 & 0.11 &  9.37 \\
 K02534.02 &  11.91 & 0.84 & 0.08 & 15.49 \\
 K02403.01 &  17.98 & 0.79 & 0.04 & 29.89 \\
 K00988.01 &  60.03 & 0.97 & 0.04 & 28.79 \\